\documentclass[12pt,preprint,authoryear]{elsarticle}
\usepackage[american]{babel}
\usepackage[latin1]{inputenc}
\usepackage{amsmath,amssymb,epsfig,natbib,bm,url,rotating,multirow,amsfonts, amscd, bezier, amstext, amsthm}
\usepackage[a4paper, text={16cm,21cm},top=2.5cm, left=2.5cm, right=2.5cm, includeheadfoot, headheight=1cm]{geometry}
\usepackage{longtable,booktabs,color}
\usepackage{epstopdf}

\title{Classification methods applied to credit  scoring:  A systematic review and overall comparison}
\author{Francisco Louzada$^a$ Anderson Ara$^b$ Guilherme B. Fernandes$^c$\\ ~\\
\small $^a$ Department of Applied Mathematics \& Statistics, University of S\~ ao Paulo, S\~ ao Carlos, Brazil \\
\small $^b$ Department of Statistics, Federal University of S\~ ao Carlos, S\~ ao Carlos, Brazil \\
\small $^c$ P\&D e Inovation in Analytics, Serasa-Experian, S\~ ao Paulo, Brazil \\
}
\date{}

\begin{document}

\begin{abstract} The  need  for controlling  and  effectively managing  credit  risk  has  led financial  institutions to  excel in improving  techniques  designed  for this  purpose,  resulting  in  the  development of various  quantitative models  by  financial  institutions and  consulting  companies.    Hence,  the  growing  number  of academic studies  about  credit  scoring shows a variety  of classification  methods  applied  to discriminate good and bad borrowers.  This paper,  therefore,  aims to present a systematic literature review relating theory and application of binary  classification  techniques  for credit  scoring financial analysis.  The general results show the use and importance of the main techniques for credit  rating,  as well as some of the  scientific paradigm changes throughout the years.


\end{abstract}

\maketitle

\section{Introduction}

The need for credit analysis was born in the beginnings of commerce in conjunction with the borrowing and lending of money, and the purchasing authorisation to pay any debt in future.    However, the modern concepts  and  ideas of credit  scoring analysis  emerged  about  70 years  ago with  \citet{Durand41}.     Since then, traders have begun to gather  information on the applicants for credit  and catalog  them  to decide between lend or not certain  amount of money \citep{paper6,paper47,paper140}.

According  to  \citet{Thomas02} credit  scoring is "a  set  of decision  models and  their  underlying techniques  that aid credit  lenders  in the  granting of credit".   A broader  definition  is considered  in the present work:  credit  scoring is a numerical  expression  based  on a level analysis  of customer  credit  worthiness,  a helpful tool  for assessment and prevention of default  risk, an important method  in credit  risk evaluation, and an active  research  area in financial risk management.

At the same time,  the modern  statistical and data  mining techniques  have given a significant contribution  to the  field of information science and  are capable  of building models to measure  the  risk level of a single customer  conditioned  to his characteristics, and then classify him as a good or a bad payer according  to his risk level.  Thus,  the  main  idea of credit  scoring models is to identify  the  features  that influence the  payment or the  non-payment behaviour  of the  costumer  as well as his default  risk, occurring the classification  into two distinct groups characterised by the decision on the acceptance or rejection  of the credit  application \citep{han2006data}.

Since the  Basel Committee on Banking  Supervision  released  the Basel Accords, specially the  second accord from 2004, the use of credit  scoring has grown considerably,  not only for credit  granting decisions but  also for risk management purposes.  The internal rating  based approaches allows the institutions to use internal ratings  to determine the risk parameters and therefore,  to calculate  the economic capital  of a portfolio Basel III, released in 2013, renders  more accurate  calculations of default  risk, especially in the consideration of external  rating  agencies, which should have periodic,  rigorous and formal comments  that are  independent  of the  business  lines under  review and  that reevaluates its  methodologies  and  models and any significant changes made to them  \citep{Rohit13,staff2013}.

Hence, the need for an effective risk management has meant that financial institutions began to seek a continuous  improvement of the techniques  used for credit analysis, a fact that resulted  in the development and application of numerous  quantitative models in this scenario.  However, the chosen technique  is often related  to the subjectivity of the analyst  or state  of the art  methods.  There are also other properties that usually differ, such as the number  of datasets applied  to verify the quality of performance  capability or even other  validation and misclassification  cost procedures.  These are natural events,  since credit scoring has been widely used in different fields, including  propositions of new methods  or comparisons  between different techniques  used for prediction  purposes  and classification.  

A remarkable, large and essential literature review was presented in the paper by \citet{paper3}, which discuss important issues of classification methods applied to credit scoring. Other literature reviews were also conducted but only focused on some types of classification methods and discussion of the methodologies, namely \citet{paper69}, \citet{paper90}, \citet{paper96} and \citet{papern62}. Also, \citet{papern42} performed a systematic literature review, but limiting the study to papers published between 2000 and 2013, these authors provided a short experimental framework comparing only four credit scoring methods. \citet{papern8} in their review considered 50 papers published between 2000 and 2014 and provided a comparison of several classifications methods in credit scoring. However, it is known that there are several different methods that may be applied for binary classification and they may be encompassed by their general methodological nature and can be seem as modifications of others usual existing methods. For instance, linear discriminant analysis has the same general methodological nature of quadratic discriminant analysis. In this sense, even though \citet{papern8} considered several classification methods they did not consider general methodologies as genetic and fuzzy methods.

This paper,  therefore,  we aim to present a more general systematic literature review over the  application of binary classification  techniques  for credit scoring,  which  features  a  better understanding of the  practical applications of credit  rating  and  its changes over time. In the present literature review, we aim to cover more than 20 years of researching (1992-2015)  including 187 papers, more than any literature review already carried out so far, completely covering this partially documented period in different papers. Furthermore, we present a primary experimental simulation study under nine general methodologies, namely, neural networks, support vector machine, linear regression, decision trees, logistic regression, fuzzy logic, genetic programming, discriminant analysis and Bayesian networks, considering balanced and unbalanced databases based on three retail credit scoring datasets.
We intent to summarise researching findings and obtain useful guidance for researchers interested in applying binary classification techniques for credit scoring.

The remainder of this paper is structured as follows. In Section \ref{content}  we present the conceptual classification scheme for the systematic  literature review, displaying  some important practical aspects  of the  credit  scoring techniques. The main credit scoring techniques are briefly presented in Section \ref{SecMAIN}. In Section \ref{results} we present the  results  of the systematic review  under  the  eligible reviewed papers,  as well as the systematic review  over four different time periods based on a historical economic context.  
In Section 5 we  compare all presented methods on a replication based study. 
Final comments  in Section \ref{final} end the paper. 

\section{Survey methodology} \label{content}

Systematic review, also known as systematic review, is an adequate alternative for identifying  and classifying key scientific  contributions to  a  field on  a  systematic, qualitative and  quantitative  description of the content in the  literature.  Interested readers  can refer to \citet{hachicha2012survey}  for more  details  on systematic literature review.   It  consists  on an  observational research  method  used  to systematically evaluate  the content of a recorded  communication \citep{kolbe1991}.

Overall,  the  procedure  for conducting a systematic review  is based  on the  definition  of sources  and procedures  for the search of papers to be analysed,  as well as on the definition of instrumental categories for the  classification  of the  selected  papers,  here based  on four categories  to understand the  historical application of the credit  scoring techniques:  year of publication, title  of the journal  where the paper  was published,  name  of the  co-authors, and  conceptual scheme based  on 12 questions  to be answered  under each published  paper.  For this  purpose,  there  is a need for defining the  criteria  to select credit  scoring papers   in the research  scope. Thus,  two selection criteria  are used in this paper  to select papers  related to the credit  scoring area to be included  in the study:

\begin{itemize}
\item The  study  is limited  to  the  published  literature available  on the  following databases:   Sciencedirect,  Engineering  Information, Reaxys  and  Scopus,  covering  20,500 titles from  5,000 publishers worldwide.
\item	The  systematic review  restricts the  study  eligibility  to journal  papers  in English,  especially considering 'credit  scoring' as a keyword  related  to 'machine  learning',  'data  mining',  'classification' or 'statistic'  topics.   Other  publication forms such  as unpublished working  papers,  master  and  doctoral  dissertations, books, conference in proceedings,  white  papers  and  others  are not  included  in the  review.  The  survey  horizon  covers a period  of almost  two decades:  from January 1992 to December  2015.
\end{itemize}

\begin{figure}[!ht]
\begin{center}
\includegraphics[scale=0.75]{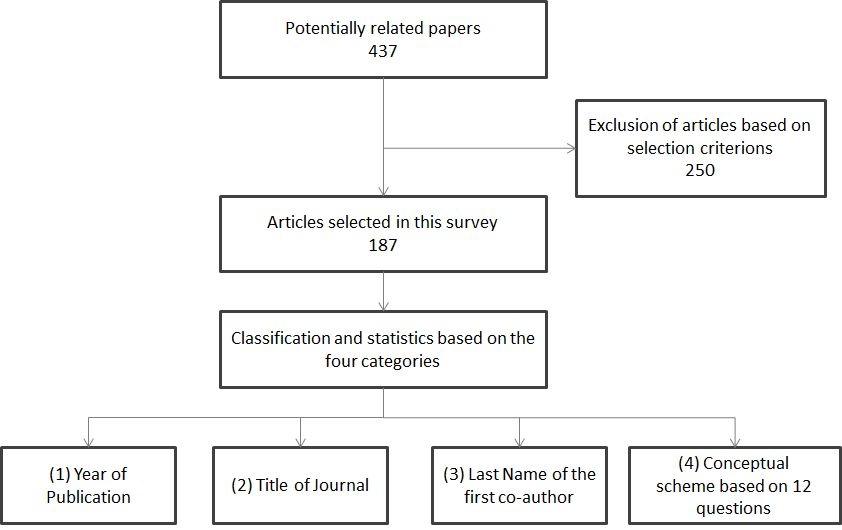} 
\par
\end{center}
\vspace{-0.5cm}
\caption{Procedure of the systematic review review.}
\label{fig_sur}
\end{figure}

The papers  were selected according  to the procedure  shown in Figure 1. From 437 papers  eligible as potentially related  to credit scoring, 250 were discarded  due to not meeting the second selection criterion. The 187 papers included in the study were  subjected  to the systematic review, according to 12 questions  on the  conceptual scenario over the  techniques:
What is the main objective of the paper? What is the type of the main classification method? Which type the datasets used?
Which is the type of the explanatory variables?
Does the paper perform variable selection methods?
Was missing values imputation performed? 
 What is the number of datasets used in the paper?
Was performed exhaustive simulation study?
What is the type of  validation of  the approach?
What is the type of  misclassification cost criterion?
Does the paper use the Australian or the German datasets?
Which is the principal classification method used in comparison study?
The  12 questions  and  possible answers  are shown in Table A.1 in the Appendix.

\subsection{The main  objective of the papers} \label{techquines}

Although  a  series  of papers is focused  on  the  same  area,  they  have  different  specific objectives. One can separate them  in general  similar  aims.   In the  present  work,  we consider  seven types  of main objectives:  proposing a new method  for rating,  comparing  traditional techniques, conceptual discussions, feature  selection,  literature review, performance  measures  studies  and,  at  last,  other  issues.  Conceptual discussions  account for papers  that deal with  problems  or details  of the  credit  rating  analysis.  In other issues, were included papers  that presented low frequency objectives.

In the proposition  of new methods,   \citet{paper17} introduce  a discriminant neural model to perform credit  rating,  \citet{paper43} propose  a support vector  machine  model within  a Bayesian  evidence framework.  \citet{paper48} propose a boosted  genetic fuzzy model, \citet{paper89} using a combined  method  that covers neural  networks,  support vector  machine  and bayesian  networks.

\citet{paper90} performed a systematic literature review that covers multiple  criteria  linear  programming models applied  to credit  scoring from 1969 to 2010.  Other  literature reviews were performed  by \citet{paper3,paper13,paper69,paper90,paper96,paper133}.

Among  the  papers  that perform  a conceptual discussion,  \citet{paper4} presents  tools  used by  the Banque  de France,   \citet{paper6}  discusses  how hazard  models  could  be considered  in order  to investigate when the borrowers  will default,  \citet{paper11} discusses the applications and challenges in credit  scoring analysis. \citet{paper95} performs  an application in credit  scoring and  discusses how their tool fits into a global Basel II credit risk management system.  Other  examples about  conceptual discussion  may  be seen in \citet{paper20}, \citet{paper47} and \citet{paper86}.  

In comparison  of traditional  techniques,  \citet{paper7} compared  five neural  network  model  with  traditional techniques. The  results  indicated  that neural  network  can improve  the  credit  scoring accuracy  and  also that logistic regression  is a good alternative to the  neural  networks.   \citet{paper21} performed  a comparison involving  discriminant analysis,  logistic  regression,  logic programing, support vector  machines,  neural networks,  bayesian  networks,  decision trees  and  k-nearest  neighbor.   The  authors concluded  that many classification techniques  yield performances which are quite competitive with each other.  Other important comparisons  may be seen in \citet{paper10,paper19,paper31,paper33,paper37,paper38,paper44,paper46,paper50,paper54,paper55,paper63,paper64,paper65,paper76,paper77,
paper81,paper83,paper97,paper108}.   Also, \citet{paper35,paper36,paper71,paper98,paper102,paper123,paper125,paper137}  handled features selection.   \citet{paper3,paper13,paper69,paper90,paper96,paper133} produced  their  work in literature review.  \citet{paper24,paper34,paper42,paper66}  worked  in performance  measures.   There  are  other  papers covering model selection  \citep{paper2}, sample  impact  \citep{paper32}, interval  credit  \citep{paper111}, segmentation and accuracy  \citep{paper130}.

\subsection{The main  peculiarities of the credit scoring papers} \label{peculiarities}

Overall the main classification methods in credit scoring are neural  networks  (NN) \citep{RIPLEY}, support  vector  machine  (SVM) \citep{vapnik1998statistical}, linear regression (LR) \citep{paper18}, decision trees (TREES)  \citep{breiman1984classification}, logistic regression (LG) \citep{berkson1944application}, fuzzy logic (FUZZY)
\citep{zadeh1965fuzzy}, genetic programming \citep{koza1992genetic}, discriminant analysis (DA) \citep{FISHER}, Bayesian networks (BN) \citep{FRIEDMAN}, hybrid methods (HYBRID)  \citep{paper17}, and ensemble methods  (COMBINED), such as bagging \citep{breiman1996}, boosting \citep{schapire1990}, and stacking \citep{wolpert1992}. 

In comparison  studies, the principal  classification  methods  involve traditional techniques  considered  by the  authors to contrast the  predictive  capability of their  proposed methodologies.   However,  hybrid  and  ensemble methods  are seldom used in comparison  studies because they involve a combination of other traditional methods.

The main  classification  methods in credit  scoring are briefly presented in the Section \ref{SecMAIN} as well as other issues related to credit scoring modeling,  such as, 
types of the  datasets used in the papers (public or not public),the use of the so called Australian or German datasets,
type  of the  explanatory variables,  
feature selection  methods, 
missing values imputation  \citep{little2002statistical}
number  of datasets used,
exhaustive simulations, 
validation approach, such as holdout sample, K-fold, leave one out, trainng/validation/test, 
misclassification cost criterions, such as Receiver Operating Characteristic (ROC) curve, metrics based on confusion matrix, accuracy (ACC), sensitivity (SEN), specificity (SPE), 
precision (PRE), false Positive Rate (FPR), and other  traditional measures used in credit scoring analysis are F-Measure  and two-sample K-S value. 

\section{The main  classification  methods in credit  scoring} \label{SecMAIN}

In this  section,  the main  techniques  used  in credit  scoring and  their applications are briefly explained  and discussed.

{\it Neural  networks  (NN).} 
A neural  network  \citep{RIPLEY} is a system  based  on input  variables,  also known  as explanatory variables,  combined  by  linear  and  non-linear  interactions through one or more hidden  layers,  resulting  in the  output variables,  also called  response  variables.    Neural  networks  were created  in an attempt to simulate  the human  brain,  since it is based on sending electronic signals between a huge number  of neurons.   The  NN structure have elements  which receive an amount  of stimuli  (the input  variables),  create  synapses  in several  neurons  (activation of neurons  in hidden  layers),  and  results in responses  (output variables).  Neural  networks  differ according  to  their  basic  structure.  In general, they  differ in the  number  of hidden  layers  and  the  activation functions  applied  to  them. \citet{paper7} shows the  mixture-of-experts and  radial  basis  function  neural  network  models  must consider for credit  scoring models. \citet{paper17} proposed  a two-stage hybrid  modeling procedure  to integrate the discriminant analysis approach with artificial neural networks technique. More recently,  different artificial neural  networks  have been suggested  to tackle  the  credit  scoring problem:  probabilistic neural  network \citep{paper28}, partial logistic artificial neural network \citep{paper78}, artificial metaplasticity neural network  \citep{paper117} and hybrid neural networks  \citep{paper113}. In some datasets, the  neural  networks  have  the  highest  average  correct  classification  rate  when compared  with other traditional techniques, such as discriminant analysis and logistic regression, taking  into account the fact that results  were very close \citep{paper64}. { Possible particular methods of neural networks are feedforward neural network, multi-layer perceptron, modular neural networks, radial basis function neural networks and self-organizing network.}

{\it Support  vector  machine  (SVM). } 
This  technique  is a statistical classification  method  and  introduced by \citet{vapnik1998statistical}.  Given a training set $\left\{ (x_{i},y_{i})\right\} $, with $i=\left\{1, \ldots, n \right\} $, where $x_{i}$ is the explanatory variable vector,  and  $y_{i}$  represents the  binary  category  of interest, and  $n$  denotes  the  number  of dimensions  of input  vectors.  SVM attempts to find an optimal  hyper-plane, making it a non-probabilistic binary  linear classifier.  The optimal  hyper-plane could be written as follows:
\[
 \sum_{i=1}^{n}w_{i}x_{i}+b=0,
\]
where  $\mathbf{w}=w_{1},w_{1}, \ldots, w_{n} $    is the  normal  of the  hyper-plane, and  $b$ is a scalar  threshold.   Considering the hyper-plane separable  with respect  to $y_{i}\in\left\{ -1,1\right\} $ and with geometric distance $\frac{2}{\left\Vert \mathbf{w}\right\Vert^2 }$, the procedure maximizes  this distance,  subject  to the constraint $y_{i}\left(\sum_{i=1}^{n}w_{i}x_{i}+b\right)\geq 1$. Commonly,  this maximization may  be  done  through the  Lagrange  multipliers   and  using  linear,  polynomial,   Gaussian  or  sigmoidal separations.  Just  recently  support  vector  machine  was considered  a credit  scoring model \citep{paper74}.  \citet{paper41,paper43,paper45,paper56,paper67,paper82,paper87,paper104,paper138,paper142} used  support vector  machine  as main technique  for their  new method. { Possible particular methods of SVM are radial basis function least squares support vector machine, linear least-squares support vector machine, radial basis function, support vector machine and linear support vector machine. 
}

{\it Linear regression (LR).}
The linear regression analysis has been used in credit scoring applications even though  the response variable  is a two class problem.  The technique  sets a linear relationship between the characteristics of borrowers  $X=\{X{1},...,X_{p}\}$ and the target variable  $Y$,  as follows,
\[
Y=\beta_{0}+\beta_{1}X_{1}+\beta_{2}X_{2}+\ldots+\beta_{p}X_{p}+\epsilon,
\]
where $\epsilon$  is the  random  error  and  independent of $X$ .  Ordinary least  squares  is the  traditional procedure to estimate  $\beta=\beta_{0},\ldots,\beta_{p}$, being $\hat{\beta}$ the estimated vector.   Once $Y$  is a binary  variable,  the  conditional expectation $E(Y|X)=x'\beta$  may be used to segregate  good borrowers  and  bad  borrowers.   Since $-\infty < x'\beta < \infty $, the  output of the  model cannot  be interpreted as a probability.  \citet{paper18}  built a superscorecard model  based  on linear  regression.   \citet{paper49}  propose  the  Poisson mixture  models for analyzing  the  credit-scoring  behaviour  for individual  loans.  Other  authors have been working with linear regression models or its generalizations in credit scoring \citep{paper18,paper22,paper49,paper93}.

{\it Decision trees (TREES).}
Classification  and Regression  Trees  \citep{breiman1984classification} is a classification method  which  uses historical  data  to  construct so-called decision  rules  organized    into  tree-like  architectures.    In general,  the  purpose  of this  method  is to determine a set  of if-then  logical conditions  that permit  prediction  or classification  of cases.  There  are three usual tree's algorithms: chi-square  automatic interaction detector (CHAID),  classification  and  regression  tree  (CART) and  C5,  which  differ by  the criterion  of tree  construction, CART  uses gini as the  splitting  criterion,  C5 uses entropy, while CHAID uses  the chi-square  test  \citep{paper122}. \citet{paper1} exhibit  a rule based model implementation in a stock selection.  \citet{paper130} used CHAID and CART  to verify the segmentation value in the  performance  capability.  \citet{paper149} proposes  a combination of a Bayesian  behavior  scoring model and a CART-based credit  scoring model. { Other possible  and particular methods of decision trees are C4.5 decision trees algorithm and J4.8 decision trees algorithm. 
}

{\it Logistic regression (LG).}
Proposed by \citet{berkson1944application}, the logit model considers a group of explanatory variables $X=\{X_1,...,X_{p}\}$ and
a response variable with two categories $Y=\{y_{1},y_{2}\}$, the technique of logistic regression consists in the estimation
of a linear combination between $X$ and the logit transformation of $Y$. Thus, if we consider $y_{1}$ as the category of interest
for analysis, the model can be represented as $log\left(\frac{\pi}{1-\pi}\right)=X\beta$,
where $\pi=P(Y=y_{1})$ and $\beta$ is the vector containing the
model's coefficients. Alternatively, the model can be represented by,
\begin{equation}
\pi_{i}=\frac{\exp\{X_{i}\beta\}}{1-\exp\{X_{i}\beta\}}, \label{eq:LR}
\end{equation}
where $\pi_{i}$ is the probability of the $i^{th}$ individual to
belong to category $y_{1}$, conditioned to $X_{i}$. The logistic regression model is a traditional method, often compared with other techniques \citep{paper15,paper34,paper29,paper64,paper122,paper148} or it is used in technique combinations \citep{paper121}. { Other possible  and particular methods of logistc regression are regularized logistic regression and limited logistic regression. 
}

{\it Fuzzy logic (FUZZY).}
\citet{zadeh1965fuzzy} introduced the Fuzzy Logic as a mathematical system which deals with  modeling  imprecise  information in the  form of linguistic  terms,  providing  an approximate answer to  a matter based  on knowledge  that is inaccurate, incomplete  or not  completely  reliable.   Unlike  the binary  logic, fuzzy logic uses the  notion  of membership  to  handle  the  imprecise  information.  A fuzzy set is uniquely determined by its membership function, which can be triangular, trapezoidal, Gaussian, polynomial or sigmoidal function.  \citet{paper19} performed an evaluation of two fuzzy classifiers for credit scoring.  \citet{paper51} proposes  a method  of building  credit  scoring models using fuzzy rule based classifiers.  \citet{paper85}  investigated the usage of Takagi-Sugeno (TS)  and  Mamdani  fuzzy models in credit  scoring. { Possible  methods in fuzzy logic  are regularized adaptive network based Fuzzy inference systems and fuzzy Adaptive Resonance. 
}

\begin{figure}[!ht]
\begin{center}
\includegraphics[scale=0.55]{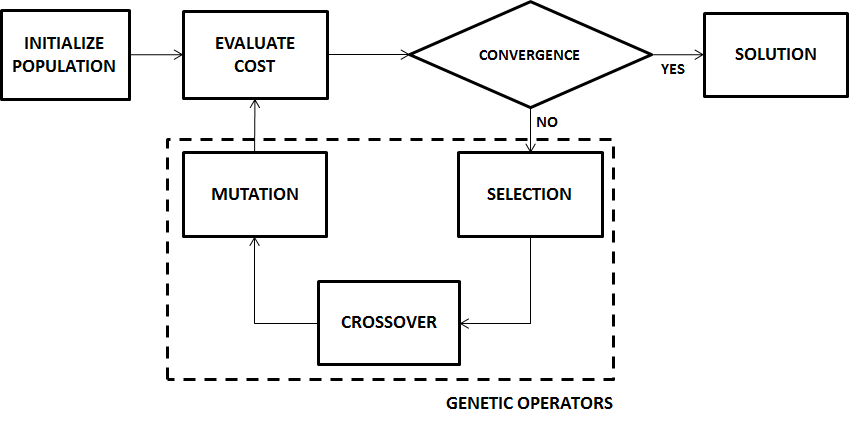} 
\par
\end{center}
\vspace{-0.70cm}
\caption{Flowchart of a genetic algorithm, adapted from \citet{abdoun2012comparative}.}
\label{fig_gen}
\end{figure}

{\it Genetic programming (GENETIC).}
Genetic Programming \citep{koza1992genetic} is based  on mathematical global optimization as adaptive heuristic search algorithms, its formulation is inspired by mechanisms of natural selection and genetics.  Basically, the main goal of a genetic algorithm is to create  a population of possible  answers  to  the  problem  and  then  submit  it  to  the  process  of evolution,  applying  genetic operations such as crossover, mutation and reproduction. The crossover is responsible for exchanging bit strings  to  generate  new observations.  Figure  \ref{fig_gen}  shows the  optimization process  of a genetic  algorithm. Ong et al. (2005) propose a genetic credit  scoring model and compares  this  with traditional techniques. \citet{paper39} introduce a two-stage genetic programming. Many other  authors have  investigated genetic models in application of credit  scoring \citep{paper20,paper23,paper81,paper150}. { Other possible  methods in genetic programming are the two stages genetic programming  and genetic algorithm knowledge refinement.
}

{\it Discriminant analysis (DA).}
Introduced by \citet{FISHER}, the discriminant analysis is based on the construction of one or more linear functions involving the explanatory variables. Consequently, the general model is given by 
\[
Z=\alpha+\beta_{1}X_{1}+\beta_{2}X_{2}+\ldots+\beta_{p}X_{p},
\]
where $Z$ represents the discrimination score, $\alpha$ the intercept,
$\beta_{i}$ represents the coefficient responsible for the linear contribution
of the $i^{th}$ explanatory variable $X_{i}$, where $i=1,2,\ldots,p$.

This technique has the following assumptions: (1) the covariance matrices
of each classification subset are equal. (2) Each classification group
follows a multivariate normal distribution. Frequently, the linear discriminant analysis is compared with other credit scoring techniques \citep{paper7,paper43,paper146} or is subject of studies of new procedures to improve its accuracy \citep{paper56,paper98}. 
{ Other possible method in discriminant analysis is quadratic discriminant analysis.
}

{\it Bayesian networks (BN).}
A Bayesian  classifier \citep{FRIEDMAN} is based  on calculating  a posterior  probability of each observation belongs to a specific class.  In other  words,  it  finds the  posterior probability distribution $P(Y|X)$, where $Y=(y_{1},y_{2},...,y_{k})$ is a random  variable  to be classified featuring $k$ categories,  and  $X=(X_{1},X_{2},...X_{p})$ is a set of  $p$  explanatory variables.   A Bayesian  classifier may  be seen as a Bayesian  network  (BNs):  a directed  acyclic graph  (DAG)  represented by the triplet ($\mathbb{V}$, $\mathbb{E}$, $\mathbb{P}$), where $\mathbb{V}$ are  the  nodes,  $\mathbb{E}$ are  the  edges and  $\mathbb{P}$ is a set  of probability distributions and  its parameters. In this  case, the  nodes represent the  domain  variables  and  edges the  relations  between  these  variables. \citet{paper8}  presents a conditional Bayesian independence graph to extract insightful information on the variables association structure in credit scoring applications. \citet{paper13}  applied bayesian networks in a credit database of annual reports  of Czech engineering enterprises. Other authors that have investigated Bayesian  nets in credit  scoring models are \citet{paper16,paper75,paper119}. { Possible methods in Bayesian networks are naive Bayes, tree augmented naive Bayes and gaussian naive Bayes.
}

{\it Hybrid methods (HYBRID).}
Hybrid methods combine different techniques to improve the performance capability.  In general,  this  combination can be accomplished  in several  ways during  the  credit  scoring process.  \citet{paper17} proposed a hybrid method that integrates the backpropagation neural networks with traditional discriminant analysis to evaluate credit scoring.    \citet{paper52}  proposed  a hybrid  method  that integrates genetic algorithm and support vector machine to perform feature  selection and  model parameters optimisation simultaneously, as well as \citet{paper17,paper29,paper30,paper40,paper68,paper74,paper107,paper120,paper134,paper139,paper146,paper148} also work with hybrid methods.

{\it Ensemble methods  (COMBINED).}
The ensemble procedure  refers to methods  of combining classifiers, thereby   multiple   techniques   are  applied  to  solve  the  same  problem  in  order  to  boost  credit  scoring performance.   There  are three  popular  ensemble  methods:   bagging \citep{breiman1996}, boosting \citep{schapire1990}, and stacking \citep{wolpert1992}.   The  Hybrid  methods  can  be  regarded  as  a  particular case  of stacking,  but  in this  paper  we consider as stacking  only the  methods  which use this  terminology.   \citet{paper127}  proposed a combined bagging decision tree to reduce the influences of the noise data  and the redundant attributes of data.  Many other authors have chosen to deal with combined methods  \citet{paper48,paper89,paper92,paper106,paper112,paper121,paper129,paper143} in credit  scoring problems.

\subsection{Other issues related to credit scoring modeling} 

{\it Types of the  datasets used. } 
As much as nowadays  the  information is considered  easy to access, mainly because  of the  modernization of the  web and  large data  storage  centers,  the  availability of data  on the credit  history  of customers  and  businesses  is still  difficult  to access.   Datasets  which  contain confidential  information on  applicants cannot  be  released  to  third  parties  without careful  safeguards \citep{Hand2001a}. Not rarely,  public data  sets are used for the investigation of techniques  and methodologies of credit rating.  In this sense, the type of dataset used (public or not public) in the papers is an important issue.

{\it Type  of the  explanatory variables. }  
The  explanatory variables,  often  known  as covariates, predictor attributes, features,  predictor  variables  or independent variables,  usually  guide the  choice and  use of a classification  method.   In general,  the  type of each explanatory variable  may be continuous  (interval or ratio)  or categorical  (nominal,  dichotomous  or ordinal).  A common practice  is to discretize  a continuous attribute as done by \citet{paper13,paper23,paper31,paper119}.  In this paper,  we consider  a continuous  dataset to be the  one that contains  only interval  or ratio  explanatory variables  - independent if a discretization method is applied or not, and a categorical dataset presents only categorical explanatory variables.  A mixed dataset is composed by both types of variable.

{\it Feature selection  methods. }  
When  we use data  to try  to provide  a credit  rating,  we use the  number of variables  that, in short,  explain  and  predict  the  credit  risk. Some methods  provide  a more accurate classification,  discarding  irrelevant features.  Thus, it is a common practice to apply  such methods  when one proposes  a  rating  model.   Some authors used  a  variable  selection  procedure  in their  papers such as \citet{paper17, paper32, paper64, paper102, paper143}.  Authors, who did not  cite or discuss feature  selection  methods  in their  papers, were regarded  as nonusers.

{\it Missing values imputation. }
The presence of missing values in datasets is a recurrent statistical problem in several application areas.  In credit  analysis,  internal  records  may be incomplete  for many  reasons:  a registration poorly conducted, the  customers  can fail to answer to questions,  or the  database or recording  mechanisms  can malfunction.  One  possible  approach is to  drop  the  missing  values  from the  original dataset, as done by \citet{paper10, paper26, paper150} or perform a preprocessing  to  replace  the  missing values,  as done  by \citet{paper22, paper33, paper92}.  These procedures are known as missing data  imputation \citep{little2002statistical}.

{\it Number  of datasets used. }  
In general, authors must  prove  the  efficiency of their  rating  methods  on either  real  or simulated datasets.  However,  due  to  the  difficulty  of obtaining  data  in the  credit  area, many times the number  of datasets used can be small, or even engage the use of other  real datasets that prove the efficiency of the rating  method.   \citet{paper42} used 16 popular  datasets in the experiments that testify  performance  measures  and which was applied  in a credit  card application.

{\it Exhaustive simulations. }
The  exhaustive simulations  study  are based  on monte  carlo sample  replications and the statistical comparisons  to assess the performance  of the estimation procedure.  In this sense, artificial samples with specific properties are randomly  generated. \citet{paper2, paper5, paper22, paper140, paper144}  are some examples of authors who performed  exhaustive simulations  in credit  scoring analysis.

{\it Validation approach. }
Amongst the various  validation procedures  we point out:
\vspace{-0.15cm}
\begin{description}
\item [{}]~~~~ {\it Holdout sample}. This validation method involves a random  partition of the dataset into two subsets: the first, called training set is designed to be used in the model estimation phase.  The second, called test set,  is used to perform  the  evaluation of the  respective  model.    Therefore, the  model is fitted  based on the training set aimed to predict the cases in the test set.  A good performance  in the second dataset indicates  that the model is able to generalize the data,  in other words, there is no overfitting on the training set.

\item [{}]~~~~ {\it K-fold.} This method  is a generalization of the hold out method,  meanwhile the data  set is randomly partitioned into K subsets.  Each subset  is used as a test  set for the model fit considering  the other K-1 subsets as training set.  In this approach, the entire dataset is used for both training and testing the model.  Typically,  a value of K=10  is used in the literature \citep{mitchell1997machine}.

\item [{}]~~~~ {\it Leave One Out}. This method is an instance  of K-fold where K is equal to the size of the dataset. Each case is used as a test set for the model fit considering  the other  cases as training set.   In this approach, the entire dataset is used for both training and testing models.   It is worth  to mention  that on large datasets a computational difficulty may arise.

\item [{}]~~~~ {\it Train/validation/test}. This  validation approach is an  alternative of the  holdout  case  for large datasets, the  purpose  is to avoid  some overfitting  into  the  validation set.   The  training samples  are used  to  develop  models,  the  validation samples  are  used  to  estimate  the  prediction  error  for the model selection,  the  test  set is used to evaluate  the  generalization error  of the  final model chosen. For this,  the performance  of the selected model should be confirmed through the measuring  of the performance  on a third  independent dataset denominated test  set \citep{bishop1995neural}.  A common split is 50\% for training, 25\% each for validation and test.

\end{description}

{\it Misclassification cost criterions.} Amongst the various  misclassification  criterions  we point out:
\vspace{-0.15cm}
\begin{description}
\item [{}]~~~~ {\it ROC curve}. The  Receiver Operating Characteristic curve was introduced by \citet{Zweig}  and  may  be geometrically  defined  as a graph  for visualizing  the  performance  of a binary classifier technique.  The  ROC  curve  is obtained by measuring  the  1?specificity on the first axis and measured  the  sensitivity on the second axis, creating  a plane.  Therefore, the  more the  curve distances from the  main  diagonal,  the  better is the  model performance.   Figure  \ref{fig_roc} shows an example  of ROC Curve.

\begin{figure}[!ht]
\begin{center}
\includegraphics[scale=0.9]{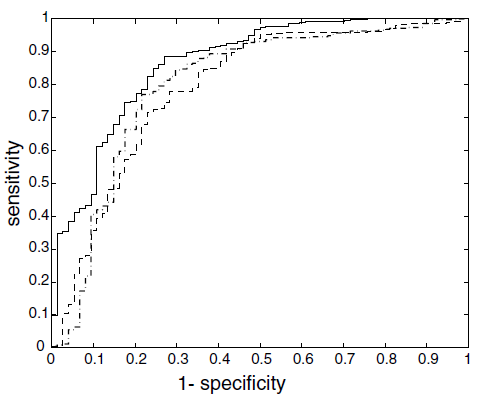} 
\par
\end{center}
\vspace{-0.5cm}
\caption{The Receiver operating characteristic curves used by \citet{paper43} to compare 
support vector machine (solid line), logistic regression (dashed-dotted line) and linear discriminant analysis (dashed line).}
\label{fig_roc}
\end{figure}

\item [{}]~~~~ {\it Metrics based on confusion matrix}.
Its  aim  is to  compare  the  model's predictive  outcome  with the true  response values in the dataset. A misclassification  takes place when the modeling procedure fails to  correctly  allocate  an  individual  into  the  correct  category.      A traditional procedure  is to build a confusion matrix, as shown in Table \ref{MC}, where $M$ is the model prediction, $D$ is the real values in data  set,  $TP$  the  number  of true  positives,  F P  the  number  of false positives,  $FN$  the  number of false negatives  and  $TN$  the  number  of true  negatives.   Naturally, $TP + FP + FN + TN  = N$ , where $N$  is the number  of observations. 
\vspace{-0.35cm}
\begin{table}[htb]
\caption{Confusion matrix.\label{MC}}
\centering{}%
\begin{tabular}{cccc}
\hline 
 &  & \multicolumn{2}{c}{$M$}\tabularnewline
 &  & \{1\} & \{0\}\tabularnewline

$D$
 & %
\begin{tabular}{c}
\{1\}\tabularnewline
\{0\}\tabularnewline
\end{tabular} & %
\begin{tabular}{c}
$TP$\tabularnewline
$FP$\tabularnewline
\end{tabular} & %
\begin{tabular}{c}
$FN$\tabularnewline
$TN$\tabularnewline
\end{tabular}\tabularnewline
\hline 
\end{tabular}
\end{table}
%\vspace{-0.35cm}

Through the  confusion  matrix, some measures  are employed  to evaluate  the  performance  on test samples.

\textit{Accuracy (ACC)}:  the  ratio  of correct  predictions of a model, when classifying cases into  class  $\{1\}$ or $\{0\}$. \textit{ACC} is defined as 
$
ACC={(TP+TN)}/{(TP+TN+FN+FP)}.
$

\textit{Sensitivity (SEN)}: also known as \textit{Recall} or \textit{True Positive Rate} is the fraction of the cases that the technique correctly classified to the class $\{1\}$ among all cases belonging to the class $\{1\}$. \textit{SEN} is defined as
$
SEN={TP}/{(TP+FN)}.
$

\textit{Specificity (SPE)}: also known as \textit{True Negative Rate} is the ratio of observations correctly classified by the model into the class $ \{0\} $ among all cases belonging to the class $ \{0\} $. \textit{SPE} is defined as
$
SPE={TN}/{(TN+FP)}.
$

\textit{Precision (PRE)}: is the fraction obtained as the number of true positives divided by the total number of instances labeled as positive. It is measured as
$
PRE=\frac{TN}/{(TN+FP)}.
$
\textit{False Negative Rate (FNR)} also known as \textit{Type I Error}  is the fraction of  $\{0\}$ cases misclassified as belonging to the  $\{1\}$ class. It is measured as
$
FNR={FN}/{(TP+FN)}.
$

\textit{False Positive Rate (FPR)} also known as \textit{Type II Error}  is the fraction of  $\{1\}$ cases misclassified as belonging to the  $\{0\}$ class. It is measured as
$
FPR={FP}/{(TN+FP)}.
$

Other  traditional measures used in credit scoring analysis are F-Measure  and two-sample K-S value. The  F-Measure  combines  both  Precision  and  Recall,  while the  K-S value  is used to  measure  the maximum  distance  between   the  distribution functions  of the  scores of the  'good payers' and  'bad payers'.

\end{description}

%%%%%%%%%%%%%%%%%%%%%%%%%%%%%%%%%%

{\it Using the  Australian and German  dataset.} 
The  Australian and  German  datasets are two public  UCI \citep{uci:2013}  datasets concerning approved  or rejected  credit  card applications. The first has  690 cases,  with  6 continuous  explanatory variables  and  8 categorical  explanatory variables.    The second has 1000 instances,  with 7 continuous  explanatory, 13 categorical  attributes. All the explanatory variables'  names and values are encrypted by symbols.  The use of these  benchmark datasets is frequent in credit  rating  papers  and  the  comparison  of the  overall  classification  performance  in both  cases is a common  practice  for the  solidification  of a  proposed  method.    \citet{paper142}  shows  an  accuracy comparison  of different authors and techniques  for Austrian and German  datasets.

{\it Principal methods  for comparison. } 
The principal  classification  methods  in comparison  studies involve traditional techniques  considered  by the  authors to contrast the  predictive  capability of their  proposed methodologies.   %These  techniques  are  the  same  exposed  in Section  \ref{techquines}. 
However,  hybrid  and  ensemble techniques  are rarely used in comparison  studies because they involve a combination of other traditional methods.

%%%%%%%%%%%%%%%%%%%%%%%%%%%%%%%%%%%%%%%%%%%%%%%%%%%%%%%%%%%%%
\section{Results and discussion} \label{results}

In this  section we present the  general results  of the  reviewed papers.  We discuss the  classification  of papers  according  to the year of publication, scientific journal,  author and conceptual scenery.  Moreover, we present a more detailed  analysis  and  discussion  of the  systematic review  for each time period,  I, II, III and IV.

\subsection{General results} 

\begin{figure}[!ht]
\begin{center}
\includegraphics[scale=0.50]{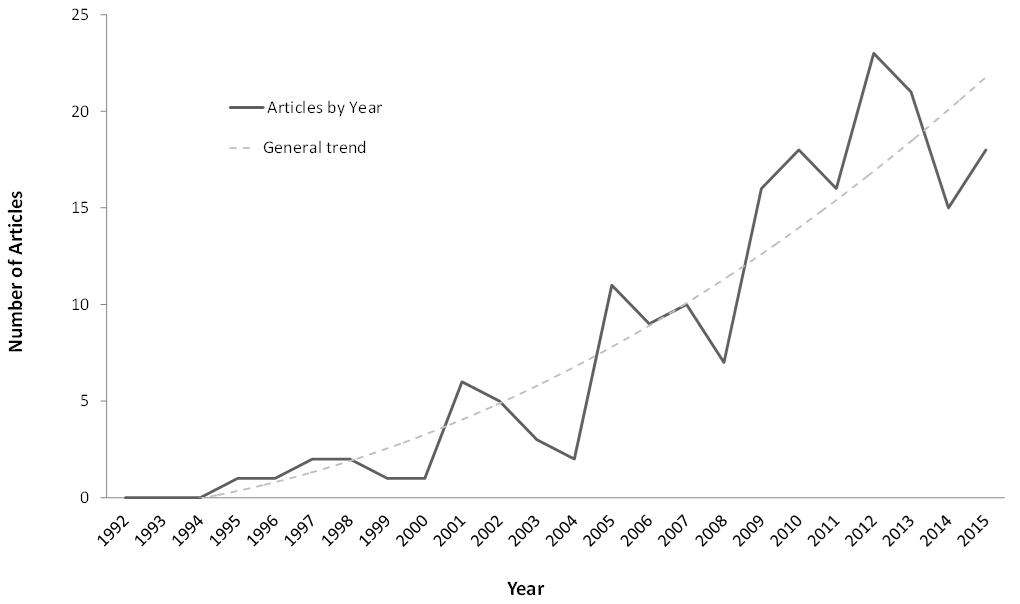} 
\par
\end{center}
\vspace{-0.5cm}
\caption{Number of credit scoring papers published by year.}
\label{papers_year}
\end{figure}

{\it On the  classification  of papers  according  to  the  year  of publication. } As indicated  in Figure  \ref{papers_year}, the number  of papers  published  in each  year  from  January 1992 to  December  2015 ranges  on 0 to  25 papers,  with  a evident growth  in all the  range  and  a fast  increment  after  2000, with  an average  of 7.8 and standard deviation  of 7.6 papers  by year.

In order to input  in the  analysis  the  historical  occurrence   we divide the  studied period of time in four  parts. The historical economic context expressed by Basel I, II and III encounters - 1988, 2004 and 2013, respectively -   may have had an increase the number  of papers  with possible time  lag in reviewing  and  revising  the  submitted manuscripts.   Thus,  we consider  the  following four  time  period  sceneries.   The  first scenery is obtained by considering  papers  published  before 2006 (Year $\le 2006$), hereafter  'I';  the second scenery is obtained by the papers  published  between  2006 and  2010 ($2006 < $Year$ \le 2010$), hereafter  'II';   the third scenery is obtained by the papers  published  between  2010 and  2010 ($2010 < $Year$ \le 2012$), hereafter  'III'; and  the last  time  scenery  is for papers  published  after  2012 (Year$>2012$) referred  to 'IV'.   The  respective  number  of papers  in each time  period  scenery equals to 45, 51, 39 and 54 papers.

{\it On the  classification of papers  according  to the  scientific journal.}   The  reviewed papers  were published  by 73 journals    and  the  frequencies  are  shown  in Table  \ref{tab:paper_journal}.  Most  of these  papers  are  related  to scientific  journals  of computer science,  decision  sciences,  engineering  and  mathematics.  As shown  in Table  \ref{tab:paper_journal}, the largest  number  of papers  were published by 'Expert  Systems with Applications'  and 'Journal of the Operational Research  Society' which account for 27.81\% and 10.70\% of the 187 reviewed paper, respectively.  In the four time periods, the journal  'Expert  Systems with Applications' exhibits  moderately an increasing number of papers in credit scoring, while the 'Journal  of the Operational Research Society' exhibits  a decreasing  number  of papers  in the same context. 'Knowledge-Based Systems' amounts to an exponential increase of these papers.

\begin{table}[!ht]
  \centering
  \caption{Distribution of reviewed papers according to the journal title in the four time periods.}
  {\footnotesize
  
\begin{tabular}{lccccccc}
\toprule
   Journal 																		&          I &         II &        III &         IV & &      Total &         \%  \\
\hline
Expert Systems with Applications 							&          9 &         16 &         14 &         13 & &         52 &    27.81 \\

Journal of the Operational Research Society 	&         11 &          5 &          1 &          3 & &         20 &    10.70 \\

European Journal of Operational Research 			&          1 &          6 &          3 &          6 & &         16 &     8.56 \\

Knowledge-Based Systems 											&          0 &          1 &          2 &          4 & &          7 &     3.74 \\

Applied Stochastic Models in Business and Industry &     4 &          0 &          0 &          0 & &          4 &     2.14 \\

Computational Statistics and Data Analysis		&          1 &          0 &          1 &          1 & &          3 &     1.60 \\

IMA Journal Management Mathematics 						&          2 &          1 &          0 &          0 & &          3 &     1.60 \\

International Journal of Neural Systems 			&          0 &          0 &          3 &          0 & &          3 &     1.60 \\

    Others 																		&         15 &         22 &         15 &         27 & &         79 &    42.25 \\
\hline
     Total 																		&         43 &         51 &         39 &         54 & &        187 &      100 \\
    \bottomrule
    \multicolumn{8}{l}{\tiny $^\dag$ These include papers from ACM Trans. on Knowledge Discovery from Data, Decision Support Systems,  Journal of the Royal Stat. Society, }\\
    \multicolumn{8}{l}{\tiny Inter. Journal of  Comp. Intelligence \& Applications, Applied Math. \&  Comp., Applied Soft Computing, Comm. in Statistics, Comp. Statistics, }\\ 
        \multicolumn{8}{l}{\tiny Credit  and Banking and others.}\\ \
    \end{tabular}%
    }
  \label{tab:paper_journal}%
\end{table}%

{\it On  the  classification  of papers  according  to  the  authors. } 
In  the  187 reviewed  papers,  there  are 525 different co-authors.  The  frequency  of appearance of those  is presented in Table  \ref{tab:citations}, where only co-authors with  over 4 appearances are shown.  Baesens B., Vanthienen J., Hand D.J. and Thomas L.C.  are the researchers  who published  the largest  number  of papers,  which represented 3.0\%, 1.9\%, 1.5\% and  1.5\%, respectively.   As may be seen in Table  \ref{tab:citations}  these  researchers  are mostly  from Belgium,  United Kingdom, Taiwan, US, Chile and Brazil.

\begin{table}[!ht]
  \centering
  \caption{Distribution of reviewed papers according to the author/co-author in the four time periods.}
    {\footnotesize

\begin{tabular}{lrrrrrrr}
    \toprule
     AUTOR & Affliation,Country &          I &         II &        III &         IV &      Total &         \% \\
    \hline
Baesens, B. & Katholieke Univ. Leuven, Belgium &          7 &          5 &          2 &          2 &         16 &        3.0 \\

Vanthienen, J. & Katholieke Univ. Leuven, Belgium &          6 &          4 &          0 &          0 &         10 &        1.9 \\

 Hand, D.J. & Imperial College London, UK &          7 &          0 &          1 &          0 &          8 &        1.5 \\

Thomas, L.C. & University of Southampton,UK &          1 &          2 &          2 &          3 &          8 &        1.5 \\

   Mues, C. & University of Southampton,UK &          1 &          2 &          3 &          1 &          7 &        1.3 \\

Van Gestel, T. & Katholieke Univ. Leuven, Belgium &          3 &          4 &          0 &          0 &          7 &        1.3 \\

Tsai, C.-F. & Nat. Chung Cheng University, Taiwan &          0 &          3 &          0 &          3 &          6 &        1.1 \\

  Bravo, C. & Universidad de Chile,  Chile &          0 &          0 &          0 &          4 &          4 &        0.8 \\

Louzada, F. & Universidade de Sao Paulo, Brazil &          0 &          0 &          3 &          1 &          4 &        0.8 \\

    Shi, Y. & University of Nebraska Omaha, US &          0 &          3 &          0 &          1 &          4 &        0.8 \\

    Others &            &         95 &        107 &        101 &        148 &        451 &       85.9 \\
    \hline
     Total &            &        120 &        130 &        112 &        163 &        525 &      100.0 \\
    \bottomrule
\end{tabular}  
    }
  \label{tab:citations}%
\end{table}%

{\it On the  classification  of papers  according  to conceptual scene. }
The  twelve questions  applied  in the systematic review  for  all  187 reviewed  papers  are  shown  in  the Table A.2 of the Appendix. In the next  section,  the analysis  and discussion of these results  are performed,  they  allow us to understand the methodological  progress occurred  in credit  scoring analysis  on the past  two decades.
% A summary  is shown in Table  A.3 of the Supplementary Material, for the four time periods I,  II, III and IV
 
\begin{figure}[!h]
  \centering
\begin{center}
\includegraphics[scale=1]{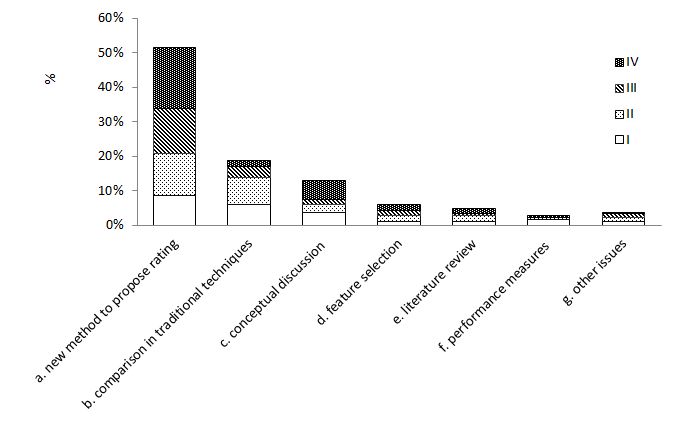} 
\par
\end{center}
\vspace{-0.9cm}
\caption{Main objectives of the credit scoring analysis.}
\label{graf.q1}
\end{figure}

%%%%%%%%%%%%%%%%%%%%%%%%%%%%%%%%%%%%%%%%%%%%%%%%%%%%%%%%%%%%% 
\subsection{Results for different time periods } \label{discussion}

\begin{figure}[!htb]
  \centering
\begin{center}
\includegraphics[scale=0.8]{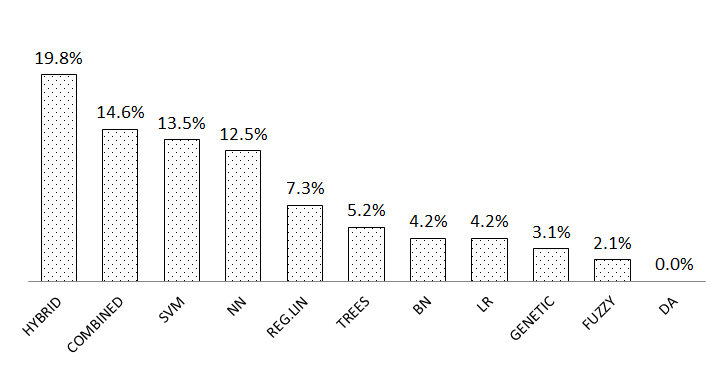} 
\par
\end{center}
\vspace{-0.5cm}
\caption{The principal techniques in proposition of new methods in credit scoring.}
\label{graf.q1-cat}
\end{figure}

{\it On the  main  objectives  in credit  scoring analysis. } 
As shown in Figure  \ref{graf.q1} the  most  common  goal of the  papers  is the  proposition  of new methods  in credit  scoring,  representing 51.3\% of all 187 reviewed papers.    This  preference  is maintained for the  four  time  periods.   Figure  \ref{graf.q1-cat} shows the  frequencies  of general  techniques  used as new methods  in credit  scoring.   The  hybrid  is the most  common method with  almost  20\%, followed by combined methods with  almost 15\% and support vector  machine along with neural  networks  with around 13\%. Due to the sheer number  of methods  involved and different kinds of behavior  in each dataset, the second most popular  main objective  is the comparison  of traditional techniques. However, it is becoming less common in the latest  years (III and IV). The third  most usual main objective  is the conceptual discussion, which is most common in IV time period.  Other  main  objectives  do not  reach  10\% of the  total  of papers  reviewed.  The  performance measures  studies  are  more  common  in past  years  (I time period).  Also, there  is stability in the four  time periods of literature review and other issues.

The research  evolution  of a new field, such as credit scoring, starts with the discovery that it is poorly investigated by researchers.  Moreover, the academic and professional interest in a particular research area is usually  boosted  by new environmental changes.  In the  case of credit  scoring, the  main environmental changes are the  rapid  increase  of storage  information and  the  processing  capacity, combined  with  the creation  of the Basel accords, which means a change in \textsl{why and how} to control credit risk.  The conceptual discussion set definitions,  ideas and problems to be faced.  The increasing number  of researchers  interested in credit  scoring culminated in the  development  and  adaptation of techniques   for tackling  the main questions.  After the techniques were developed,  methods  for comparing  those are  proposed.   At last,  a field of research  will eventually  reaches  a state-of-the-art phase,  followed by new researchers  questioning  the paradigm, ideas and disrupting the status quo of the credit  scoring area.
Currently, credit  scoring is going through the process of tools development, as shown in Figure  \ref{papers_year}. %ref{tab:desc_geral}.

\begin{figure}[!ht]
  \centering

\begin{center}
\begin{tabular}{c}
General \tabularnewline
\includegraphics[scale=0.30]{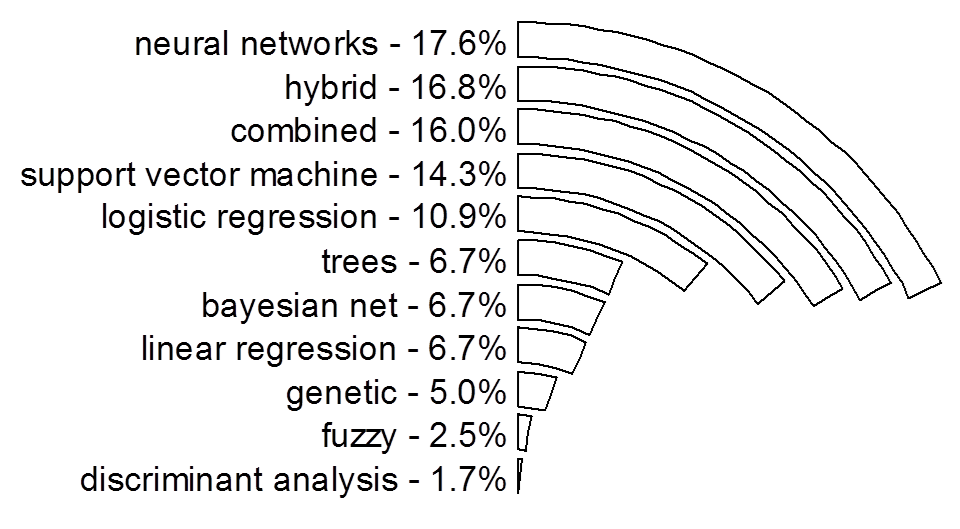} \tabularnewline

 \tabularnewline
\end{tabular}
{\small
\begin{tabular}{cc}
 I																						&  II  \tabularnewline	
\includegraphics[scale=0.30]{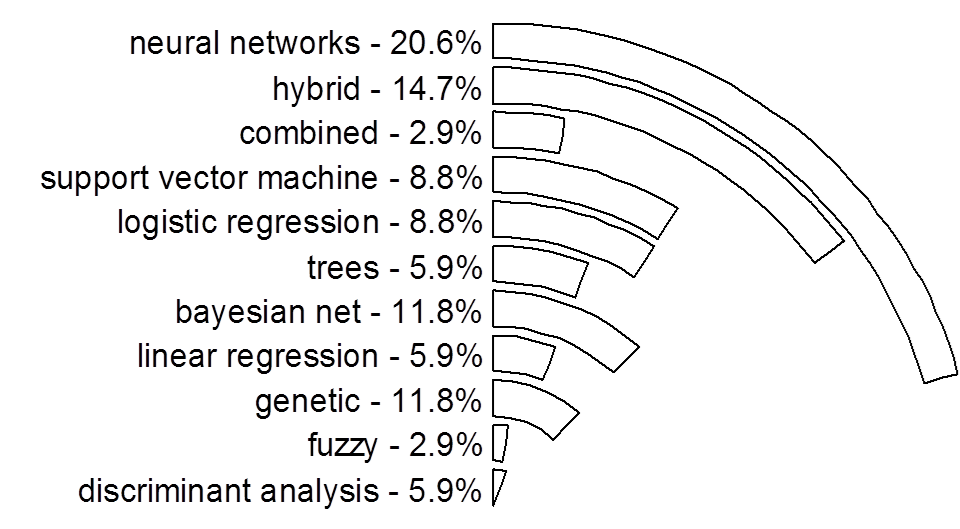}  	&  \includegraphics[scale=0.30]{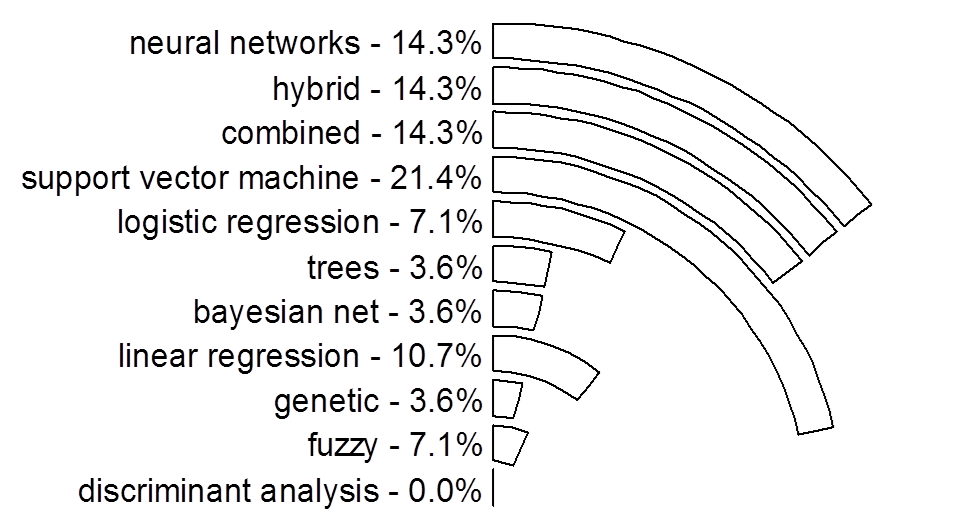} \tabularnewline				
III 																					&  IV \tabularnewline		
\includegraphics[scale=0.30]{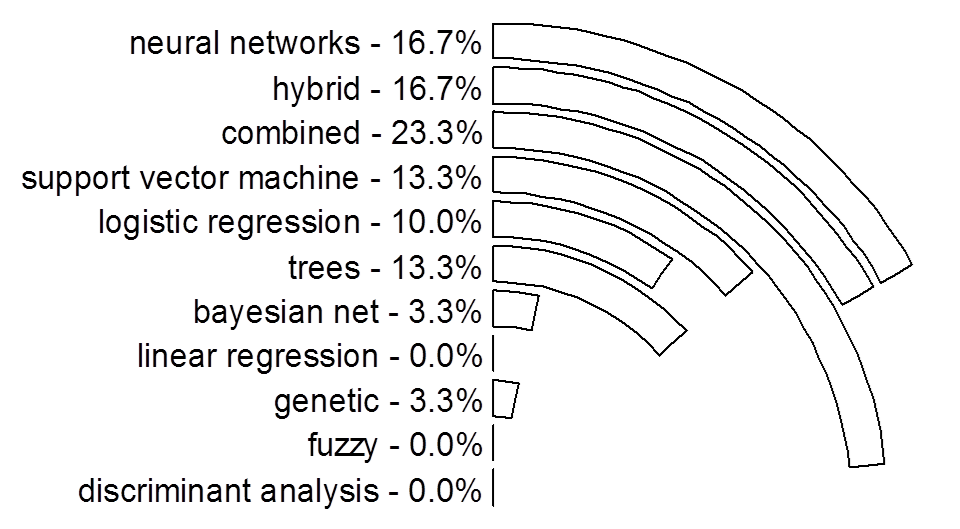}		&  \includegraphics[scale=0.30]{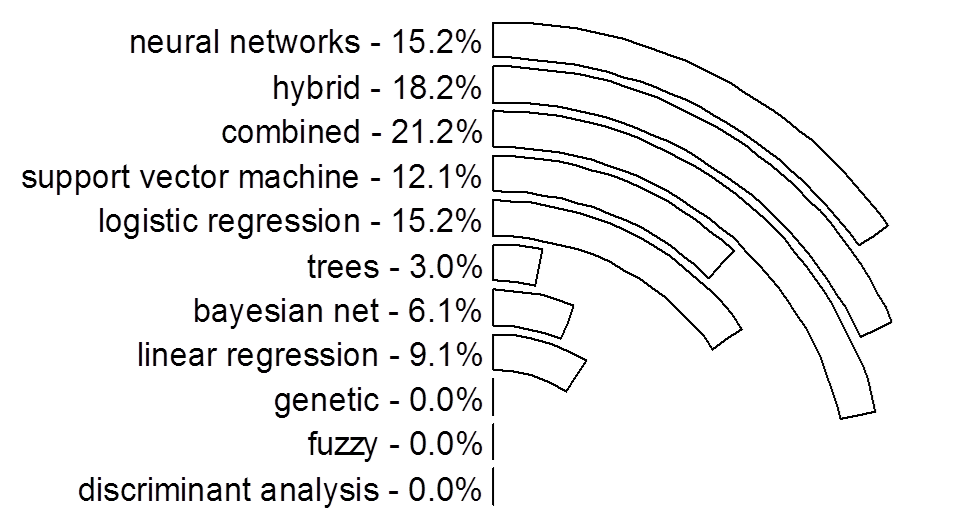} \tabularnewline

\end{tabular}
} 

\caption{The main classification techniques in credit scoring.}
\label{techs}
\end{center}
\end{figure}

{\it On  the  main  classification  techniques. }
As a classification  technique  is applied  as a credit  scoring model,  the  choice of technique  is often  related  to  the  subjectivity of the  analyst  or to  state-of-the-art methods.  Ideally,  a precise prediction  indicates  whether  a credit  extended  to an applicant will probably result  in profit  for the  lending  institution.  Figure  \ref{techs}  shows the  circular  bar  plots  concerning  the  main classification techniques  applied in all considered periods as well as their utilization over time, this Figure only  considers  the  techniques  indicated   in  Section  2.2.   In  general,  the  neural  networks  and  support vector machine are the most common used techniques  in credit scoring (17.6\%), the discriminant analysis remained  as a rarely used technique  (1.7\%).  In the first time period analyzed, the most common technique is the  neural  network  (20.6\%).

However,  neural  networks  and hybrid methods remained  at  constant use in all following periods  considered  with  a higher  frequency.  Support  vector  machine  was most  used between  2006 and 2010  II time period  (21.4\%), this  method  is the  fourth most commonly used in general, although with a fast increasing in past and decreasing  its participation over recent years.  The trees,  bayesian net, linear regression and logistic regression  techniques  had  this  same  percentage  in this  period. However, logistic regression was most used  (15.2\%) in recent years  and matching the use of neural networks in IV time period.   In addition, there is a strong decrease in the use of the genetic, fuzzy and discriminant analysis methods  and a remarkable growth  of combined  techniques  which  are  the  most  used  method  in recent  years,  IV time period,  with 21.2\%. Hybrid methods  have always been highly used, but were not the highlights  in any time period.  In comparison  with Figure \ref{graf.q1-cat}, the hybrid  and combined methods  are mostly used in new methods  to propose rating  in credit  scoring, followed by support vector  machine  and neural  networks.

\begin{figure}[!ht]

\begin{center}
\begin{tabular}{cc}
 \includegraphics[scale=0.7]{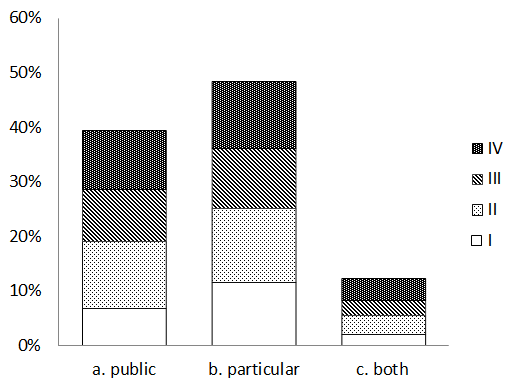}  & \includegraphics[scale=0.7]{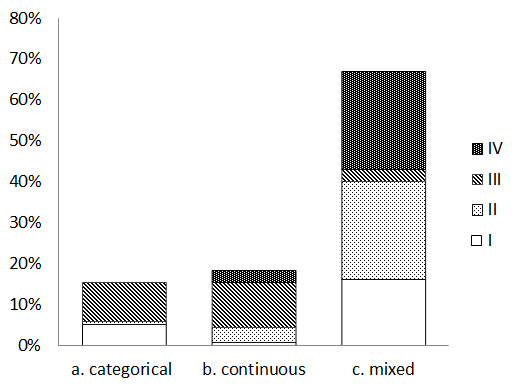} \tabularnewline
  (a)  & (b)\tabularnewline
\end{tabular}
\caption{(a) the type of used datasets and (b) the type of variables in datasets.}
\label{graf.dataset}
\end{center}
\end{figure}

\begin{figure}[!ht]
\begin{center}
\includegraphics[scale=0.5]{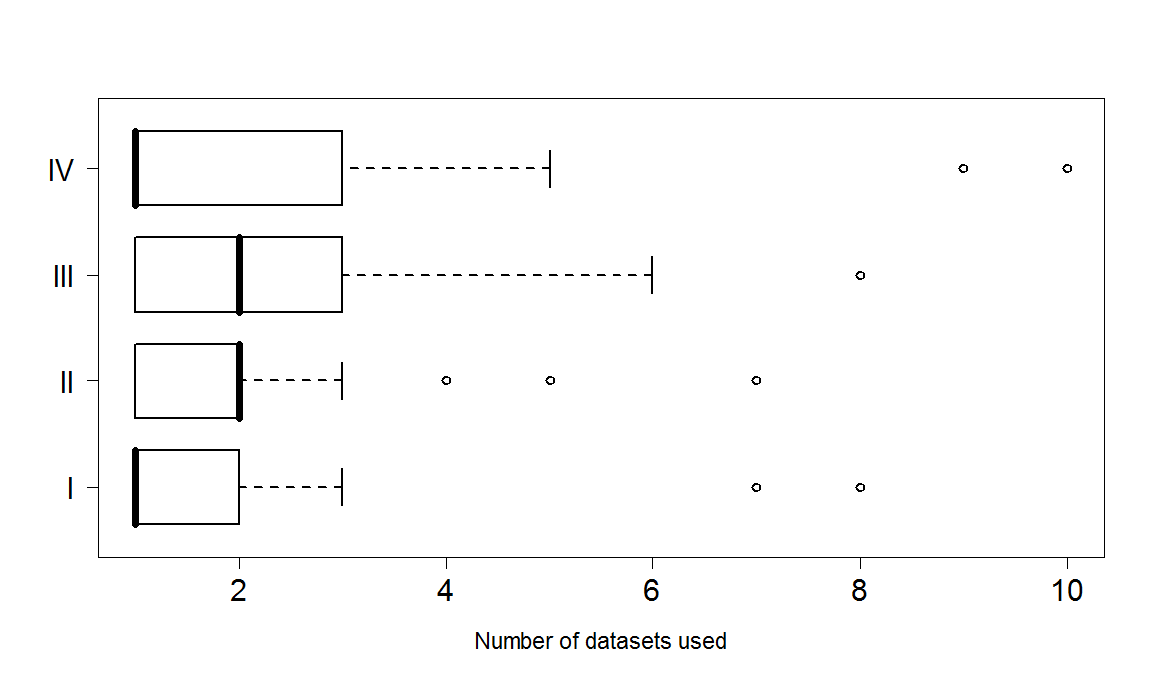} 
\par
\end{center}
\caption{The behavior of the number of dataset used in credit scoring studies.}
\label{boxplots}
\end{figure}

{\it On the  datasets used in credit  scoring. } 
Figures  \ref{graf.dataset} (a)  and  \ref{graf.dataset} (b) show information about  the  datasets used  in credit  scoring  reviewed  papers.   As Indicated by  \ref{graf.dataset} (a),  the  most common  type of datasets is private  in all time  periods,  followed by public  dataset and  lastly  the  use of both  types.  In other  words, the  authors usually  employ  only private  datasets in their  credit  scoring  applications.  This  fact  seems to be independent  of the  time  period.   As indicated  by \ref{graf.dataset} (b),  authors prefer  to use datasets that have continuous  and  discrete  variables.   However, in I the  datasets with  only discrete  variables were more common  than  those  with  only continuous  variables.   Discarding  \citet{paper42}, which used 16 datasets in their  work, Table  \ref{tab:datas_stat} shows the  basic statistics of the  number  of datasets used in reviewed papers.   In general,  the  papers  consider  an average of  2.18 datasets in their  content.  Figure  \ref{boxplots} shows the behaviour  of the  number  of datasets in the  four times  periods,  and  indicates  a growth  in the  number  of datasets used in the period I, II and III and an average decrease in IV with a growth in the standard deviation.

\begin{table}[!ht]
  \centering
  \caption{Statistical summary of the number of used datasets.}
    \begin{tabular}{lccccccc}
    \toprule
    Time period & Min. & 1st Qu. & Median & Mean & 3rd Qu. & Max. & Sdv \\
    \hline
    I	  &1.00 & 1.00 & 1.00 & 1.80 & 2.00 & 8.00 & 1.69 \\
    II 	&1.00 & 1.00 & 2.00 & 2.05 & 2.00 & 7.00 & 1.40 \\
    III &1.00 & 1.00 & 2.00 & 2.55 & 3.00 & 8.00 & 1.85 \\
    IV 	&1.00 & 1.00 & 1.00 & 2.31 & 3.00 & 10.00& 2.32 \\
    \hline
General &1.00 & 1.00 & 1.00 & 2.18 & 3.00 & 10.00 & 1.84 \\
    \bottomrule
    \end{tabular}%
  \label{tab:datas_stat}%
\end{table}%

\begin{figure}[!ht]

\begin{center}
\begin{tabular}{cc}
 \includegraphics[scale=0.6]{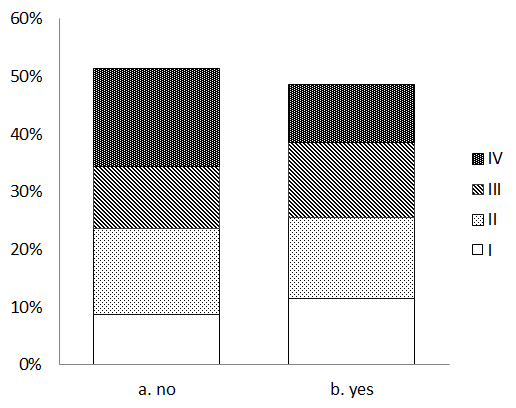}  & \includegraphics[scale=0.6]{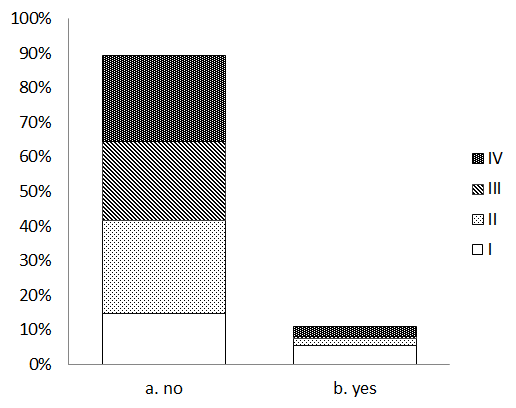} \tabularnewline
  (a)  & (b)\tabularnewline
\end{tabular}
\caption{(a) the using of feature selection and (b) the using of missing data imputation.}
\label{preprocd}
\end{center}
\end{figure}

{\it On the  preprocessing  data  methods  in credit  scoring.  }
In regards  to preprocessing  methods  in credit scoring, this review covers two relevant aspects:  the feature  (variable) selection and missing data  procedures.  Figure  \ref{preprocd}  (a)  shows that, independently of the  time  period,  the  feature  selection  is performed  in most studies.  However, in about  49\% of the papers  this procedure  is not used.  Figure  \ref{preprocd}  (b) shows that the missing data  imputation  it is a procedure often not used in credit  scoring analysis  (90\%).

\begin{figure}[!ht]
\begin{center}
\begin{tabular}{cc}
   \includegraphics[scale=0.6]{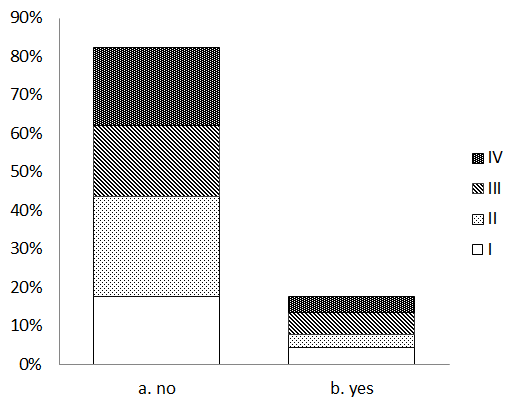}  & \includegraphics[scale=0.6]{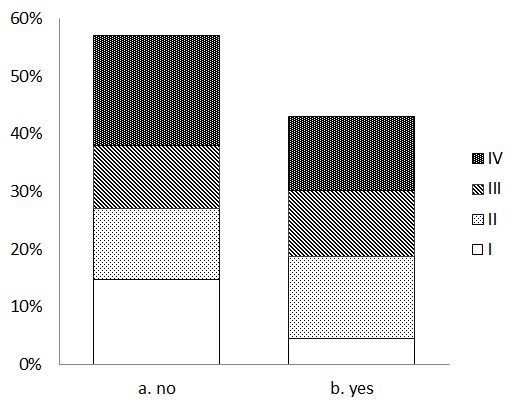} \tabularnewline
  (a)  & (b)\tabularnewline
\end{tabular}
\caption{(a) the using of exhaustive simulations and (b) the using of the Australian or German credit dataset. }
\label{others}
\end{center}
\end{figure}

\begin{figure}[!ht]
\begin{center}
\includegraphics[scale=0.95]{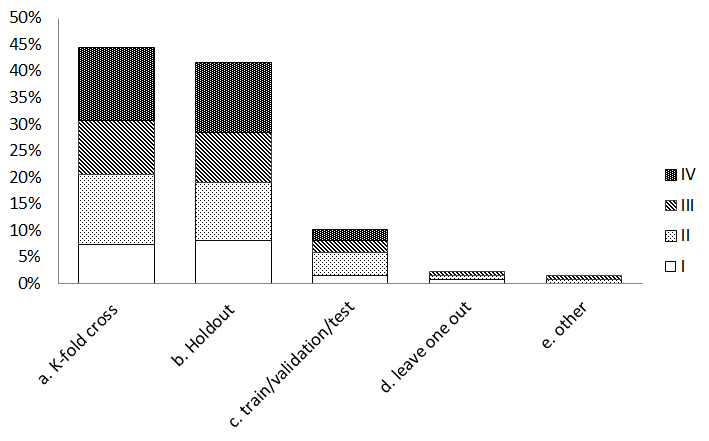} 
\par
\end{center}
\vspace{-0.5cm}
\caption{The type of validation methods.}
\label{graf.q7}
\end{figure}

{\it On the validation of the approaches.}
The validation of the approaches are a part of the procedures  that ensures the  authors of the  examination of the  performance  and  comparability of their  methods.   In general,  as indicated  by Figure \ref{others} (a), more than  80\% of the papers  do not consider exhaustive simulations  in their procedures.  Likewise, as indicated in Figure \ref{others} (b),  almost  45\% of all reviewed  papers  consider  the  Australian or German  credit  dataset, and  during the  II time  period  it  became  an  even more  common  practice.   Table    \ref{tab:paper_AUGE} shows the  overall classification performance  on Australian and German  credit datasets for 30 reviewed papers.  Concerning the splitting  of the datasets, Figure \ref{graf.q7} shows that K-Fold cross validation and holdout  methods  were more common in general, and in more recent time periods, the K-fold cross validation became the most widely used method.  The splitting  of the dataset in three parts  (train/validation/test) is more used than the leave-one-out  procedure.

\begin{table}[htbp]
  \centering
  \caption{Overall classification performance on Australian and German credit datasets.}
  {\small
  \label{tab:paper_AUGE}%
   % Table generated by Excel2LaTeX from sheet 'Plan3'
\begin{tabular}{rrrrrrr}
\toprule
     Paper &        AUS &        GER &            &      Paper &        AUS &        GER \\ 
\hline
 \citet{paper21}  &      90.40 &      74.60 &            &   \citet{paper109}  &      87.30 &      79.20 \\

 \citet{paper30}  &      98.00 &      98.50 &            &   \citet{paper117}  &      92.75 &      84.67 \\

 \citet{paper36}  &      92.60 &      83.80 &            &   \citet{paper120}  &      87.52 &      76.60 \\

 \citet{paper42}  &      86.96 &      74.40 &            &   \citet{paper123}  &      85.65 &      72.60 \\

 \citet{paper48}   &      85.80 &      73.40 &            &   \citet{paper126}   &      85.36 &      77.10 \\

  \citet{paper52}  &      87.00 &      78.10 &            &   \citet{paper127}  &      88.17 &      78.52 \\

 \citet{paper59}  &      97.32 &      78.97 &            &   \citet{paper138}  &      85.98 &      75.08 \\

 \citet{paper63}  &      90.20 &      79.11 &            &   \citet{paper139}  &      88.55 &      77.40 \\

 \citet{paper71}  &      81.93 &      74.28 &            &   \citet{paper141}  &      86.81 &      76.60 \\

 \citet{paper76}  &      86.52 &      84.80 &            &   \citet{paper142}   &      87.85 &      79.55 \\

  \citet{paper85}  &      88.60 &      75.00 &            &         \citet{papern17} &      84.83 &      73.51 \\

 \citet{paper102}  &      86.52 &      76.70 &            &         \citet{papern25} &      88.84 &      73.20 \\

 \citet{paper106}  &      91.97 &      81.64 &            &         \citet{papern31} &      86.09 &      74.16 \\

 \citet{paper107}  &      86.84 &      75.75 &            &        \citet{papern32} &      87.23 &      76.48 \\

  \citet{paper108}   &      86.57 &      76.30 &            &         \citet{papern46} &      86.78 &      76.62 \\ \bottomrule

\end{tabular}

  }
\end{table}%

\begin{figure}[!ht]
\begin{center}
\includegraphics[scale=0.95]{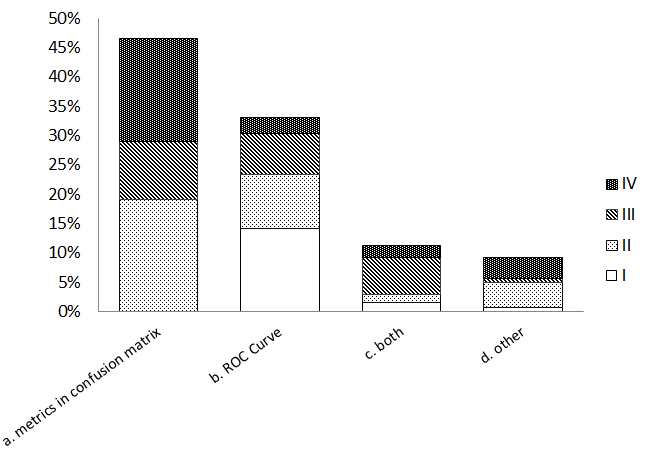} 
\par
\end{center}
\vspace{-0.5cm}
\caption{The misclassification criterions.}
\label{mcc}
\end{figure}

{\it On  the  misclassification   cost criterion.  } 
Figure \ref{mcc} shows that to  measure  the  misclassification   cost,  the  most  common criteria  used in the  reviewed  papers  are  the  metrics  based  on confusion  matrix  (45\%).   Although  this criterion  was not  in use solely in the  I time  period,  it was widely used in others. The  utilization of the  ROC  Curve  was more common in the  past  period and  about
10\% of all the reviewed papers  used both  or other criteria.

\begin{figure}[!ht]
  \centering

\begin{center}
\begin{tabular}{c}
General \tabularnewline
\includegraphics[scale=0.37]{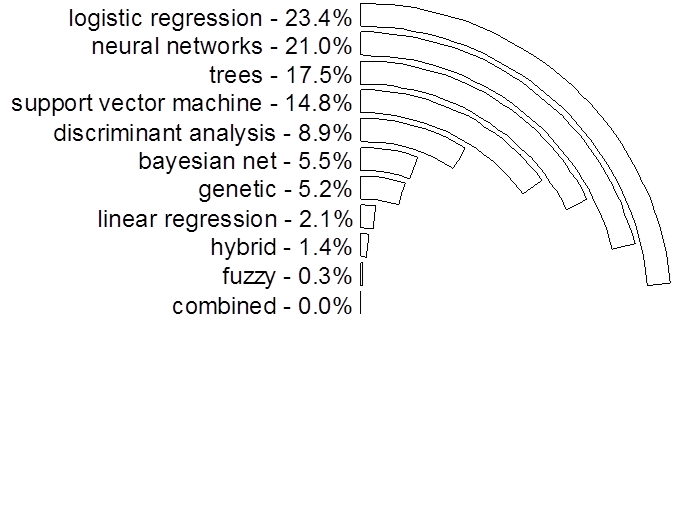} \tabularnewline

 \tabularnewline
\end{tabular}
{\small
\begin{tabular}{cc}
 I																							&  II  \tabularnewline	
\includegraphics[scale=0.37]{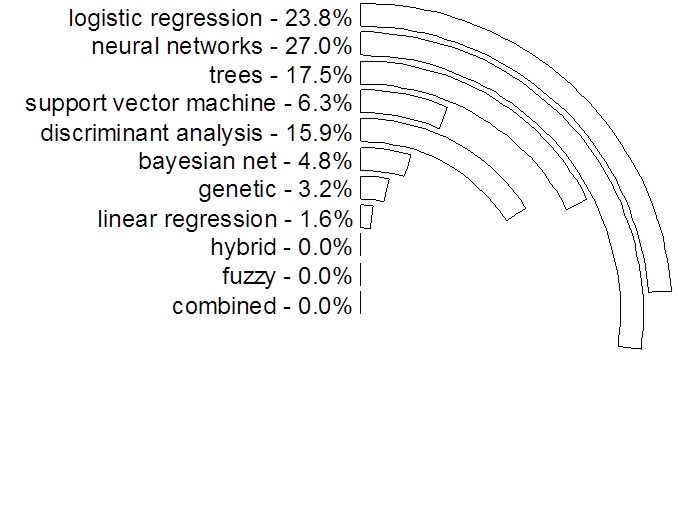}  	&  \includegraphics[scale=0.37]{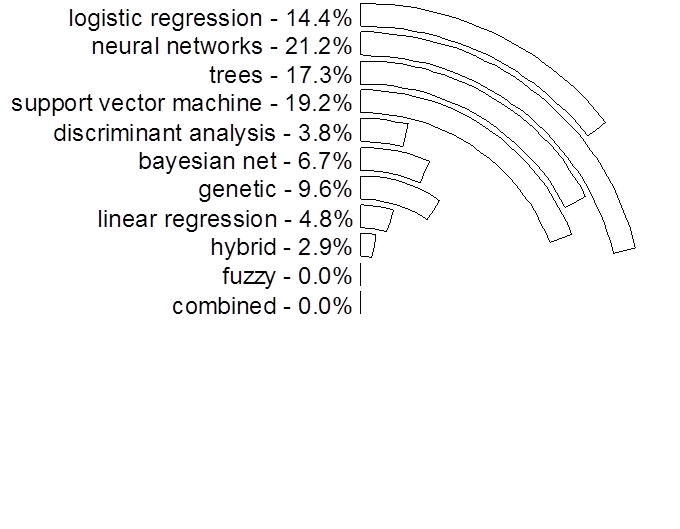} \tabularnewline				
III 																						&  IV \tabularnewline	
\includegraphics[scale=0.37]{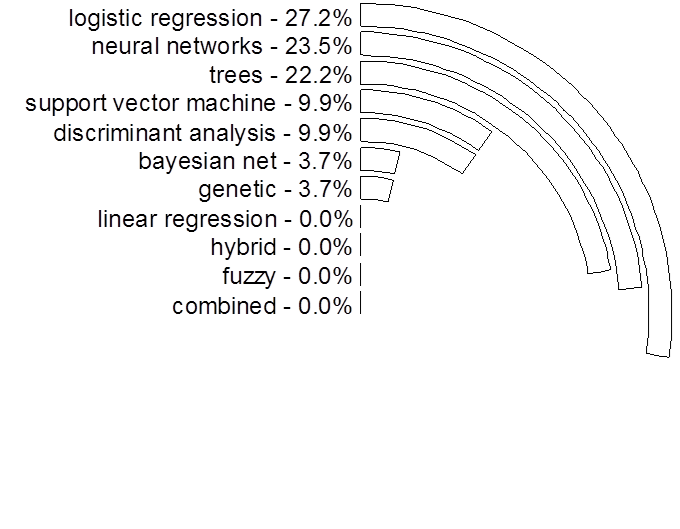}		&  \includegraphics[scale=0.37]{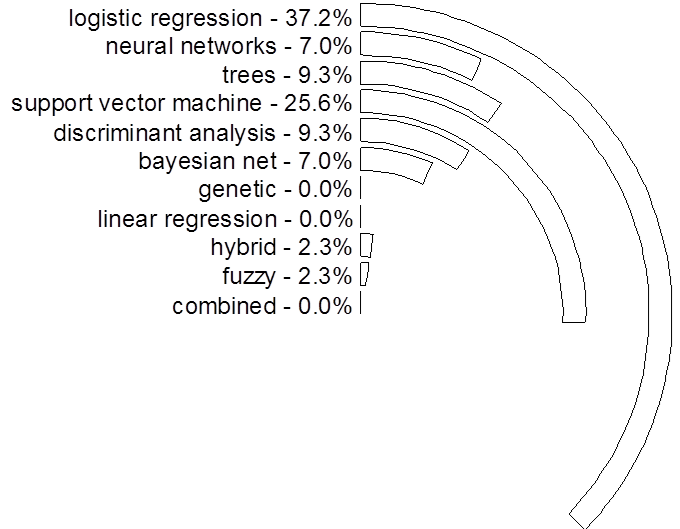} \tabularnewline

\end{tabular}
} 

\caption{The techniques used in the paper's comparison studies.}
\label{techs_comp}
\end{center}
\end{figure}

{\it On the  classification  methods  used in comparison  studies.}   Regarding  the  traditional techniques used in comparison  studies,  Figure  \ref{techs_comp} shows the circular  bar plots concerning  techniques  applied in all considered  periods.     The  most  used  technique  in comparison  studies  is logistic regression (23.14\%) which has  always  had  a high frequency  of use in all considered  periods.   The  neural networks  is the second most used technique  (21.0\%) with a high usage in II time period. The support vector  machine  was widely and recently used in comparison  studies, but  in general it is the fourth  most frequently  used technique  (14.8\%).  The trees remained  as the  third  most  used  technique  in all periods.  In the  reviewed  papers,  no study  performs comparisons  using combined techniques.

\section{Is there a better method? A comparison study} \label{guide}

In this section, all presented methods are compared using two frameworks, marked out by two predictive performance measures, AC (Approximate Correlation) and FM (F1-score Measure) for three different benchmark datasets: (A) Australian Credit, (B) Credit German and (C) Japanese Credit, available in UCI Machine Learning Repository ({\url{http://archive.ics.uci.edu/ml}/). For each dataset we performed 1000 replications in a handout validation approach (70\% training sample and 30\% test sample) under a balanced base ($p = 0.5$, 50\% of bad payers) and a unbalanced base ($p = 0.1$, 10\% of bad payers). The methods were implemented in Sofware R 3.0.2 through RBase with the packages: $nnet$, $MASS$, $rpart$, $rgp$, $e1071$ and $frbs$ on a HP Pavilion PC i7-3610QM 2.30GHz CPU, RAM 8.GB, Windows 7 64-bit.

\begin{figure}[!ht]
\begin{center}
\includegraphics[scale=0.95]{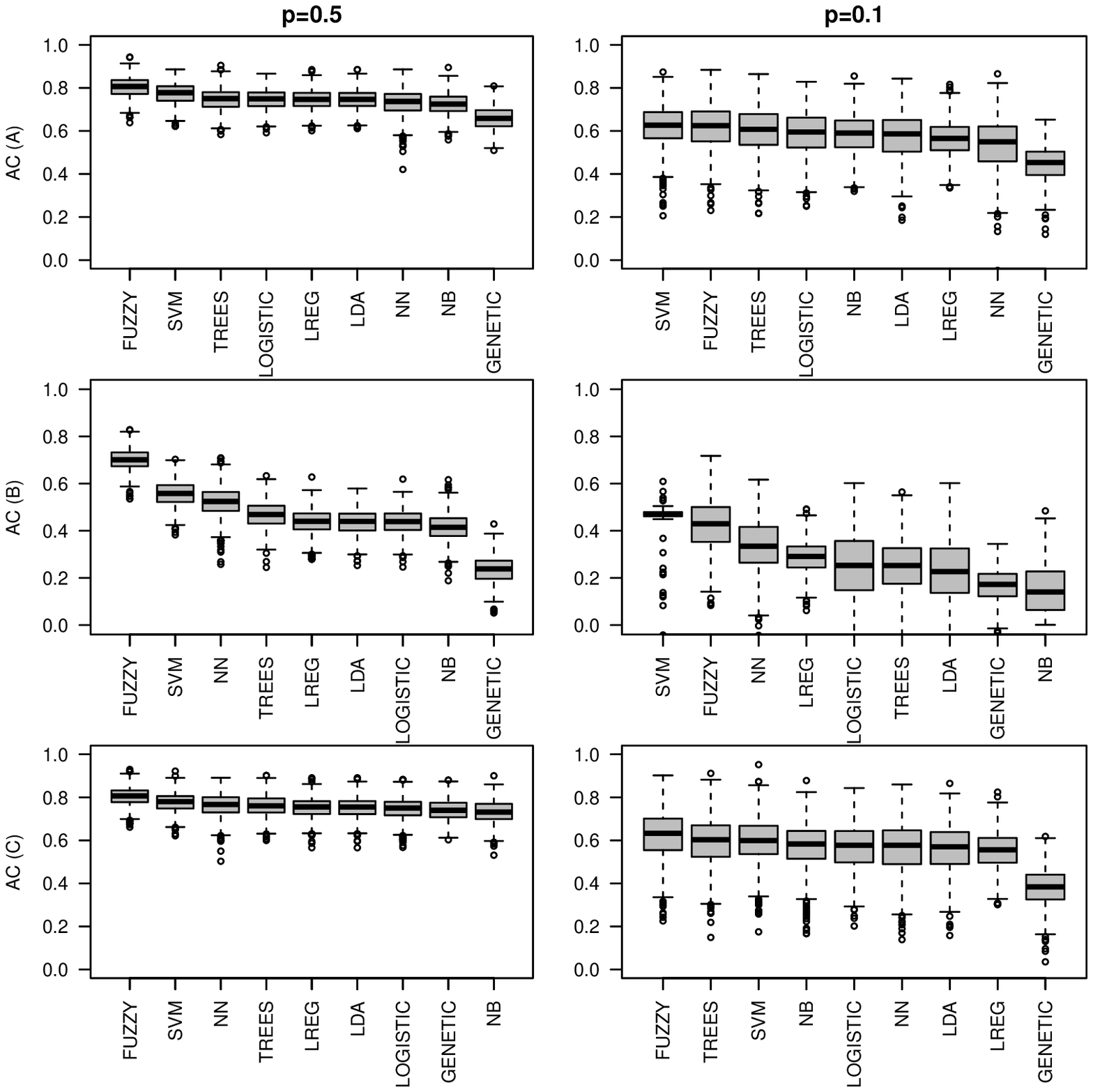} 
\par
\end{center}
\vspace{-0.5cm}
\caption{Approximate Correlation results.}
\label{AC}
\end{figure}

\begin{figure}[!ht]
\begin{center}
\includegraphics[scale=0.95]{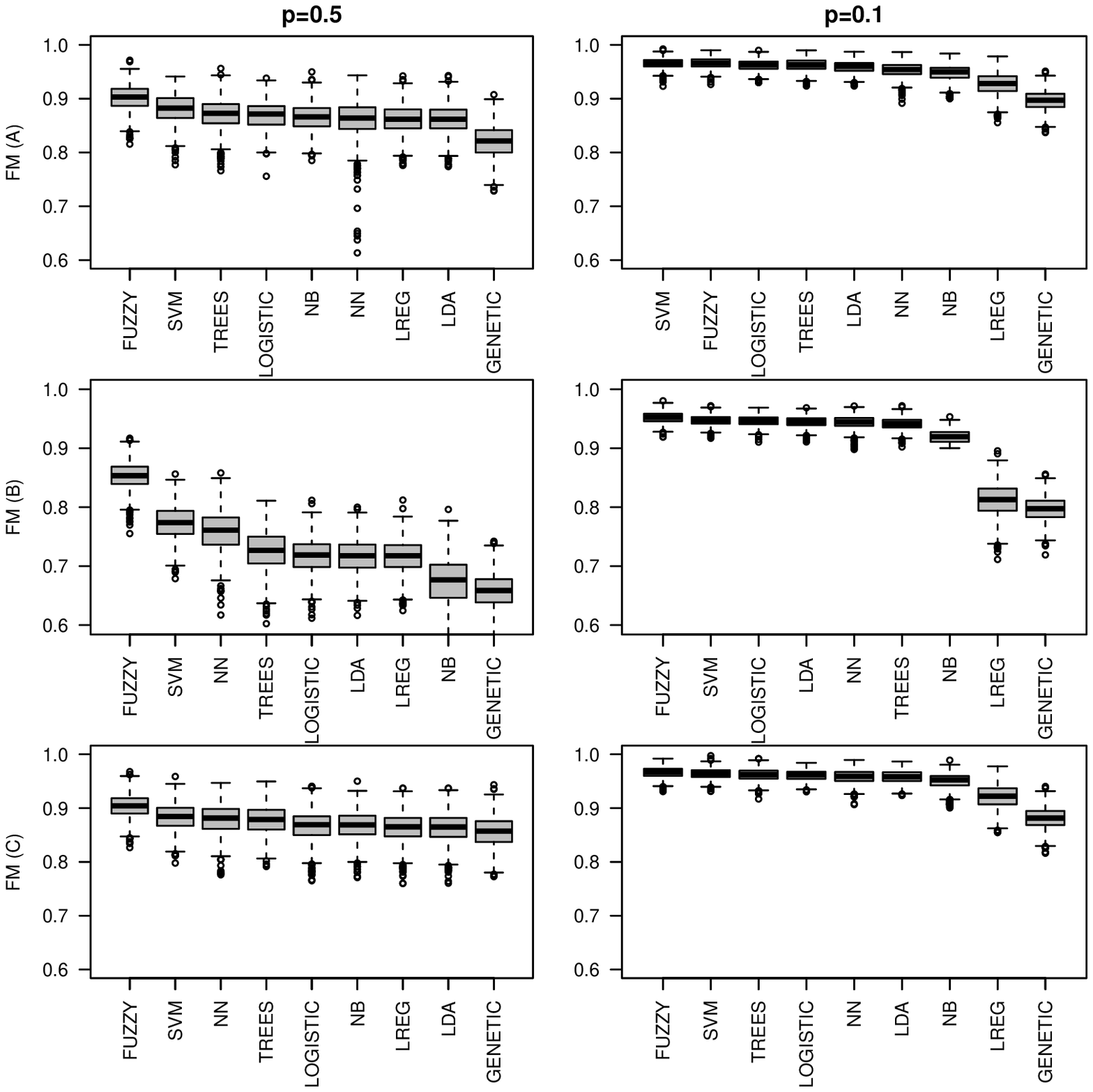} 
\par
\end{center}
\vspace{-0.5cm}
\caption{F1-score measure results.}
\label{FM}
\end{figure}

Taking into account the all comparisons, Figures \ref{AC} and \ref{FM},  we noticed the highlight of two methods, SVM and FUZZY, that permeate this comparison study as the two best techniques of greater predictive performance for both measures evaluated, this fact is confirmed by the Kruskal-Wallis test at a significance level of 5\% ($p-values <2e-16$).

However we noticed that in most cases there is a shift of the predictive performance of both when occurs unbalance in the number of bad payers. For $p = 0.5$ FUZZY is given as the method with greater predictive performance, with SVM as the second method. For $p = 0.1$ SVM is given as the method with greater predictive performance, with FUZZY as the second method. Alternatively, TREES often the third best method and its independent of the  unbalance. In addition, we noticed most often that NN lost the predictive performance when there is imbalance. The LOGISTIC, NB and LDA methods do not seem present any standard, with predictive performance behavior between the median methods. GENETIC and LREG are considered as the smaller predictive performance when there is imbalance.

% Table generated by Excel2LaTeX from sheet 'Plan1'
\begin{table}[!ht]
{\footnotesize
  \centering
  \caption{Time in seconds for each method's replication}
    \begin{tabular}{rccccccccc}
    \toprule
    {Dataset} & {GENETIC} & {FUZZY} & {NN} & {LOGISTIC} & {SVM} & {NB} & {TREES} & {LDA} & {LREG} \\
    \midrule
    (A) & 41.69 & 39.69 & 0.66 & 0.24 & 0.34 & 0.34 & 0.25 & 0.30 & 0.23 \\
    (B) & 42.58 & 24.31 & 0.41 & 0.23 & 0.31 & 0.33 & 0.27 & 0.23 & 0.23 \\
    (C)  & 146.35 & 82.77 & 1.38 & 0.74 & 0.47 & 0.45 & 0.28 & 0.25 & 0.24 \\ \midrule
    AVERAGE & 76.87 & 48.92 & 0.82 & 0.40 & 0.37 & 0.37 & 0.27 & 0.26 & 0.23 \\
    \bottomrule
    \end{tabular}%
  \label{tab:compf}%
  }
\end{table}%

Table \ref{tab:compf} displays computational time (in seconds) for the implementation of methods for each replication.
Among the methods with greater predictive performance, SVM (0.37s) has a much lower computational effort than FUZZY (48.92s). GENETIC and FUZZY are the methods with higher computational effort.
In summary among the analysed methods, SVM stands out as a method of high predictive performance and low computational effort than others. 

\section{Final comments} \label{final}

We present  in this  paper  a methodologically  structured systematic literature review of binary  classification techniques  for credit  scoring financial analysis.  An amount of 187 papers  on credit  scoring  published  in scientific  journals  during  the  two last  decades  (1992-2015) were analysed  and classified.
Based on the survey, we observed an increasing  number  of papers  in this area and noticed  that credit scoring analysis  is a current and significant financial area.  A plenteous area for the application of statistical and data  mining tools.  
{

Although, regardless of the time period, the  most common main objective of the revised papers  is to propose a new method  for  rating  in credit  scoring,  especially with  hybrid  techniques, it is observed a similarity between the predictive performance of the methods. 
This result is corroborated by \cite{Hand2006}.
Moreover, comparison with traditional techniques was rarely performed in recent time periods. This fact show that, though the researchers are giving up to compare techniques, the pursuit of a general method with a high predictive performance continues. On the other hand, other types of researches  in credit scoring are required as conceptual discussions based on data quality, database enrichment, time dependence, classes type and so on.  

While knowing these mishaps, for the moment, neural  networks, support vector  machine,  hybrid  and combined  techniques  appear  as the most common main tools. The logistic regression, trees and also neural networks are mostly used in comparisons  of techniques  as standards that must  be overcome. In general, support vector machine appears as a method of high predictive performance and low computational effort than other methods.  Regarding  datasets for credit scoring, the  number  has been increasing  as well as the  presence of a mixture of continuous  and  discrete  variables.   The  majority of datasets however are private  and  there  is a wide usage of the well known German  and Australian datasets.  This fact shows how difficult is to obtaining datasets on the credit scoring scenario, since there are issues related to maintenance  of confidentiality of credit scoring databases.

The K-fold cross validation and holdout are the most common validation methods. Care should be taken when interpreting the results of both methods, because they are different methods and subject to subjectivity of the random distribution of the database.
%This fact should be taken into consideration when there is a comparison between results of methods used by different researchers, since there is difference in split balance of the dataset and, furthermore, in both cases this split is performed randomly. 
%
The use of ROC Curve as unique misclassification criterion has decreased significantly in the articles over the years. More recently it is most common the use of metrics based on confusion matrix.
Also, there  is a small number  of papers  that handles  with  missing  data  in credit  scoring  analysis  and  a  high  frequency  of papers  that applied  feature  selection  procedures  as pre-proceeding method.
}

Although  our  systematic literature review  is exhaustive, some limitations still  persist.   First,  the  findings  were based on papers  published  in English and in scientific journals  inside the following databases:  Sciencedirect,  Engineering  Information, Reaxys  and  Scopus.   Although  such  databases covers more  than  20,000 journal  titles,  other  databases may be hereafter  included  in the survey.  Secondly, as pointed  out in Section  2, we did not include in the survey other  forms of publication such as unpublished working papers, master  and doctoral  dissertations, books, conference in proceedings,  white papers  and others.  Moreover, high  quality  research  is eventually published  in scientific  journals,  other  forms  of publication may  be included  in this list in future  investigations. Notwithstanding these limitations, our systematic review  review provides important  insights  into the research literature on classification  techniques  applied  to credit scoring and how this area has been moving over time.

\bigskip
\noindent \textbf{{\large \textbf{Acknowledgments}}}:  This research was sponsored by the Brazilian organizations CNPq and FAPESP and by Serasa Experian, through  their research grant programs. 

%%%%%%%%

\newpage
\section*{Appendix} \label{apen}
\subsection*{Table A.1}

\begin{table}[!ht]
  \centering
%\caption
  {Table A.1: List  of questions and  of possible responses to  the  proposed systematic review.}
  
{\scriptsize
    \begin{tabular}{lll}
    \toprule
    \textit{1. Which is the main objective of the paper?} &   & \textit{6. Was missing values imputation performed?} \\
 
    a. proposing a new method for rating &   & a. yes \\
    b. comparing traditional techniques &   & b. no \\
    c. conceptual discussion  &   &  \\
    d. feature selection &   & \textit{7. What is the number of datasets used in the paper?} \\
    e. literature review &   &  \\
    f. performance measures  &   & \textit{8. Was performed exhaustive simulation study?} \\
    g. other issues &   & a. yes \\
      &   & b. no \\
    \textit{2.   What is the type of the main classification method?} &   &  \\
    a. neural networks &   & \textit{9. What is the type of  validation of  the approach?} \\
    b. support vector machine &   & a. K-fold cross \\
    c. linear regression &   & b. Handout \\
    d. trees &   & c. train/validation/test \\
    e. logistic regression &   & d. leave one out \\
    f. fuzzy &   & e. other \\
    g. genetic &   &  \\
    h. discriminant analysis &   & \textit{10. What is the type of  misclassification cost criterion?} \\
    i. bayesian net &   & a. ROC curve \\
    j. hybrid &   & b. metrics based on confusion matrix \\
    k. combined &   & c. both \\
    l. other &   & d. other \\
      &   &  \\
    \textit{3. Which type the datasets used?} &   & \textit{11. Does the paper use the Australian or the German datasets?} \\
    a. public &   & a. yes \\
    b. particular &   & b. no \\
    c. both  &   &  \\
      &   & \textit{12. Which is the principal classification method used in } \\
     &   & \textit{comparison study?} \\
    \textit{4. Which is the type of the explanatory variables?} &   & a. neural networks \\
    a. categorical &   & b. support vector machine \\
    b. continuous &   & c. linear regression \\
    c. mixed &   & d. trees \\
      &   & e. logistic regression \\
    \textit{5.  Does the paper perform variable selection methods?} &   & f. fuzzy \\
    a. yes &   & g. genetic \\
    b. no &   & h. discriminant analysis \\
      &   & i. bayesian net \\
      &   & j. other \\
    \bottomrule
    \end{tabular}%
  \label{tab:surveytab}%
  
}
\end{table}%

\newpage
\subsection*{Table A.2} 

{\scriptsize
Table A.2: Indexing of 187 reviewed papers (1992-2015).
\begin{longtable}{lrrrrrrrrrrrr}

%\caption{Indexing of reviewed papers (1992-2012)} \\

\toprule
Reviewed paper & Q01 & Q02 & Q03 & Q04 & Q05 & Q06 & Q07 & Q08 & Q09 & Q10 & Q11 & Q12 \\ 
\hline

\endfirsthead
\multicolumn{13}{l}{\tablename\ \thetable\ -- \textit{Continued from previous page}} \\
\hline
Paper & Q01 & Q02 & Q03 & Q04 & Q05 & Q06 & Q07 & Q08 & Q09 & Q10 & Q11 & Q12 \\  \hline
\endhead
\hline
\multicolumn{13}{l}{\textit{Continued on next page}}\\
\endfoot
\bottomrule
\endlastfoot
     {\tiny  \citet{paper0}}  & b  &  a &    &    &    &    &    &    &    &    &    &   ae  \\
    {\tiny  \citet{paper1}}  & a  &    &    &    &    &    &    &    &    &    & b  & d  \\
    {\tiny  \citet{paper2}}  & g  &  d & b  & c  & a  & b  &  1 & b  & b  & a  & b  &   eh  \\
    {\tiny  \citet{paper3}}  & e  &    &    &    &    &    &    &    &    &    &    &    \\
    {\tiny  \citet{paper4}}  & c  &  h & b  & a  & a  & b  &  1 & b  & c  &    &    & h  \\
    {\tiny  \citet{paper5}}  & a  &  e & b  & c  & a  & b  &  1 & a  & b  &    & b  & e  \\
    {\tiny  \citet{paper6}}  & c  &  l & b  & a  & b  & b  &  1 & b  & b  & a  &    & e  \\
    {\tiny  \citet{paper7}}  & b  &  a & a  & c  & b  & b  &  2 & b  & a  &    & b  & adej  \\
    {\tiny  \citet{paper8}}  & a  &  i & a  & a  & a  & b  &  1 & a  & a  & d  & a  & i  \\
    {\tiny  \citet{paper9}}  & c  &    &    &    &    &    &    &    &    &    &    &    \\
    {\tiny  \citet{paper10}}  & b  &  l & b  & c  & b  & b  &  1 & a  &    &    & b  &    \\
    {\tiny  \citet{paper11}}  & c  &    &    &    &    &    &    &    &    &    &    &    \\
    {\tiny  \citet{paper12}}  & c  &    &    &    &    &    &    &    &    &    &    &    \\
    {\tiny  \citet{paper13}}  & e  &  i &    &    &    &    &    &    &    &    &    & i  \\
    {\tiny  \citet{paper15}}  & a  &    & b  &    & a  & b  &  1 & a  & b  & a  & b  & e  \\
    {\tiny  \citet{paper16}}  & a  &  i &    &    &    &    &    &    & b  & c  & b  & aehi  \\
    {\tiny  \citet{paper17}}  & a  &  j & b  & c  & a  & b  &  1 & b  &    & a  & b  & aehj  \\
    {\tiny  \citet{paper18}}  & a  &  c & b  & a  & b  & b  &  1 & b  &    &    & b  & c  \\
    {\tiny  \citet{paper19}}  & b  &  f & b  & c  & b  & b  &  1 & b  & b  & a  & b  &  dfg  \\
    {\tiny  \citet{paper20}}  & c  &   ag & a  & c  & a  & b  &  1 & b  & a  & a  & a  &   ag  \\ 
    {\tiny  \citet{paper21}}  & b  & hebaidl & c  & c  & b  & b  &  8 & b  & b  & c  & a  &   abdehij  \\
    {\tiny  \citet{paper22}}  & a  &  c &    & c  & a  & a  &  1 & b  & b  & c  & b  & c  \\
    {\tiny  \citet{paper23}}  & a  &   ag & a  & a  &    &    &  3 & b  &    & a  & a  &  adg  \\
    {\tiny  \citet{paper24}}  & f  &    &    &    &    &    &    &    &    & c  &    &    \\
    {\tiny  \citet{paper25}}  & c  &    &    &    &    &    &    &    &    &    &    &    \\
    {\tiny  \citet{paper26}}  & g  &    &    &    &    &    &    &    &    &    &    &    \\
    {\tiny  \citet{paper29}}  & a  &  j & b  & c  & a  & a  &  1 & b  & a  & a  & b  & aehj  \\
    {\tiny  \citet{paper30}}  & a  &  j & a  & c  & b  & b  &  2 & b  & a  & a  & a  & j  \\
    {\tiny  \citet{paper31}}  & b  &  g & a  & c  & b  & b  &  2 & b  &    &    & b  & adeg  \\
    {\tiny  \citet{paper32}}  & g  &  e & b  & c  & a  & b  &  1 & b  & b  & c  &    & e  \\
    {\tiny  \citet{paper33}}  & b  &   al & b  & c  & b  & a  &  1 & b  & b  & a  & b  &   aj  \\
    {\tiny  \citet{paper34}}  & f  &    &    &    &    &    &    &    &    &    &    &    \\
    {\tiny  \citet{paper35}}  & d  &    & b  & c  & a  & b  &  1 & b  & a  & a  & b  &   de  \\
    {\tiny  \citet{paper36}}  & d  &  l & a  & b  & a  & b  &  7 & b  &    & a  & a  &    \\
    {\tiny  \citet{paper37}}  & b  &  j & c  & c  & a  & a  &  3 & b  &    &    & b  &     abcej  \\
    {\tiny  \citet{paper38}}  & b  &  k & b  & a  & a  & a  &  1 & b  & c  & a  & b  &     adehj  \\
    {\tiny  \citet{paper39}}  & a  &  g & a  & c  & b  & b  &  2 & b  &    &    & b  &  deg  \\
    {\tiny  \citet{paper40}}  & a  &  j & a  & c  & b  & b  &  2 & b  & b  & a  & b  & a  \\
    {\tiny  \citet{paper41}}  & a  &  b & b  & c  & a  & b  &  1 & b  & a  &    & b  &   ab  \\
    {\tiny  \citet{paper42}}  & f  &  l & a  & c  & b  & b  & 16 & b  & a  & a  & a  & adhi  \\
    {\tiny  \citet{paper43}}  & a  &  b & b  & c  & a  & a  &  1 & b  & d  & c  & b  &  beh  \\
    {\tiny  \citet{paper44}}  & b  &  a & b  & a  & b  & a  &  1 & a  & a  & a  & b  &  adi  \\
    {\tiny  \citet{paper46}}  & b  &    & c  & c  & a  & a  &  3 & a  & a  &    & b  &     bdehj  \\
    {\tiny  \citet{paper47}}  & c  &    &    &    &    &    &    &    &    &    &    &    \\
    {\tiny  \citet{paper48}}  & a  &  k & c  & c  & b  & b  &  7 & a  & b  & a  & a  &     adghi  \\
    {\tiny  \citet{paper49}}  & a  &  c & b  & c  & b  & b  &  1 &    &    & c  & b  &  bcd  \\
    {\tiny  \citet{paper50}}  & b  &    & b  & c  & a  & a  &  1 & b  & d  & b  & b  &  bce  \\
    {\tiny  \citet{paper51}}  & a  &  f & a  & c  & b  & b  &  1 & b  & b  & b  & a  & f  \\
    {\tiny  \citet{paper52}}  & a  &  j & a  & c  & a  & b  &  2 & b  & a  & b  & a  &     abdgj  \\
    {\tiny  \citet{paper53}}  & b  &    & b  & b  & a  & b  &  1 & b  & e  & a  & b  &  bde  \\
    {\tiny  \citet{paper54}}  & b  &  b & a  & c  & b  & b  &  5 & b  & b  & b  & a  & b  \\
    {\tiny  \citet{paper55}}  & b  &    & b  & b  & a  & b  &  1 & b  & a  & c  & b  &     abdei  \\
    {\tiny  \citet{paper56}}  & a  &  b & b  & b  & a  & b  &  2 & b  & a  & b  & b  &   be  \\
    {\tiny  \citet{paper57}}  & a  &  a & c  & c  & b  & b  &  3 & b  & b  & b  & a  &   ad  \\
    {\tiny  \citet{paper59}}  & a  &  a & a  & c  & b  & b  &  3 & b  & b  & b  & a  & a  \\
    {\tiny  \citet{paper60}}  & c  &    &    &    &    &    &    &    &    &    &    &    \\
    {\tiny  \citet{paper62}}  & b  &   ec & b  & c  & a  & b  &  1 & b  & c  & d  & b  &   ce  \\
    {\tiny  \citet{paper63}}  & b  &    & a  & c  & b  & b  &  4 & b  & b  & b  & a  &   ab  \\
    {\tiny  \citet{paper64}}  & b  &    & b  & c  & a  & b  &  1 & b  & b  & b  & b  &  aeh  \\
    {\tiny  \citet{paper65}}  & b  &    & b  & b  & b  & b  &  2 & b  & a  & c  & b  & deij  \\
    {\tiny  \citet{paper66}}  & f  &  e & b  &    & b  & b  &  1 & a  &    &   bd  &    & e  \\
    {\tiny  \citet{paper67}}  & a  &  b & a  & c  & b  & b  &  1 & b  & a  & b  & a  & b  \\
    {\tiny  \citet{paper68}}  & a  &  j & a  & c  & a  & b  &  1 & b  & c  & b  & a  &     adehj  \\
    {\tiny  \citet{paper69}}  & e  &    &    &    &    &    &    &    &    &    &    &    \\
    {\tiny  \citet{paper70}}  & g  &    &    &    &    &    &    &    &    &    &    &    \\
    {\tiny  \citet{paper71}}  & d  &  a & a  & c  & a  & b  &  5 & b  & a  & b  & a  & a  \\
    {\tiny  \citet{paper72}}  & a  &  d & b  & c  & a  & b  &  1 & b  & b  & b  & b  & d  \\
    {\tiny  \citet{paper73}}  & b  &    & b  & c  & b  & b  &  1 & b  & a  &   bd  & b  &    bdegij  \\
    {\tiny  \citet{paper74}}  & a  &  j & b  & c  & a  & b  &  1 & b  & b  & b  & b  &  bdj  \\
    {\tiny  \citet{paper75}}  & b  &  i & c  & c  & b  & b  &  2 & b  & b  & c  & a  & dehi  \\
    {\tiny  \citet{paper76}}  & b  &    & a  & c  & b  & b  &  2 & b  & a  & b  & a  &  bgj  \\
    {\tiny  \citet{paper77}}  & b  &    & b  & c  & a  & b  &  2 & b  & c  & b  & b  &  ceg  \\
    {\tiny  \citet{paper78}}  & a  &  a & b  & c  & b  & a  &  2 & a  & a  &    & b  & a  \\
    {\tiny  \citet{paper81}}  & b  &  g & b  & c  & a  & b  &  1 & b  & b  & b  & b  &   gj  \\
    {\tiny  \citet{paper82}}  & a  &  b & a  & c  & b  & b  &  2 & b  & a  & a  & a  & bcde  \\
    {\tiny  \citet{paper83}}  & b  &  l & b  & a  & a  & b  &  1 & b  & a  &   cd  & b  &     abdei  \\
    {\tiny  \citet{paper85}}  & a  &  f & a  & c  & a  & b  &  2 & b  & c  & b  & a  &    abdfgj  \\
    {\tiny  \citet{paper86}}  & c  &    &    &    &    &    &    &    &    &    &    &    \\
    {\tiny  \citet{paper87}}  & a  &  b & c  & c  & b  & b  &  2 & b  & c  & c  & a  &    bdegij  \\
    {\tiny  \citet{paper88}}  & g  &    &    &    &    &    &    &    &    &    &    &    \\
    {\tiny  \citet{paper89}}  & a  &  k & a  & c  & a  & b  &  1 & b  & a  & b  & a  &  abi  \\
    {\tiny  \citet{paper90}}  & e  &  l & b  &    & b  & b  &  5 & b  & a  &   bd  &    & b  \\
    {\tiny  \citet{paper92}}  & a  &  k & b  & c  & a  & a  &  1 & b  & a  & a  & b  &  abd  \\
    {\tiny  \citet{paper93}}  & a  &  c & b  &    & b  & b  &  1 & a  &    &    & b  & c  \\
    {\tiny  \citet{paper95}}  & c  &  l &    &    &    &    &  3 &    & c  &    &    &    \\
    {\tiny  \citet{paper96}}  & e  &    &    &    &    &    &    &    &    &    &    &    \\
    {\tiny  \citet{paper97}}  & b  &    & b  & c  & b  & b  &  1 & b  &   ab  & b  & b  &    \\
    {\tiny  \citet{paper98}}  & d  &  l & a  &    & a  & b  &  3 & a  & b  & a  & a  &    \\
    {\tiny  \citet{paper99}}  & a  &  l & a  & c  & b  & b  &  2 & b  &    & d  & a  &    \\
    {\tiny  \citet{paper100}}  & a  &  l & c  & b  & b  & b  &  4 & b  & a  &    & a  &  abd  \\
    {\tiny  \citet{paper102}}  & d  &    & a  & c  & a  & b  &  2 & b  & a  & c  & a  & b  \\
    {\tiny  \citet{paper104}}  & a  &  b & a  & c  & a  & b  &  1 & b  & b  & a  & b  & b  \\
    {\tiny  \citet{paper106}}  & a  &  k & a  & c  & a  & b  &  2 & b  & a  & a  & a  &  abd  \\
    {\tiny  \citet{paper107}}  & a  &  j & a  & c  & b  & b  &  2 & b  & b  & a  & a  &  bgj  \\
    {\tiny  \citet{paper108}}  & b  &  k & c  & c  & b  & b  &  3 & b  & b  & a  & a  & abde  \\
    {\tiny  \citet{paper109}}  & a  &  l & a  & c  & b  & b  &  2 & b  & a  & a  & a  &   de  \\
    {\tiny  \citet{paper110}}  & c  &    &    &    &    &    &    &    &    &    &    &    \\
    {\tiny  \citet{paper111}}  & g  &    &    &    &    &    &    & a  & b  &    & b  &    \\
    {\tiny  \citet{paper112}}  & a  &  k & b  & a  & a  & b  &  2 & b  & b  & a  & b  &  deh  \\
    {\tiny  \citet{paper113}}  & a  &  a & a  & b  & a  & b  &  1 & b  & c  & a  & a  &  aeh  \\
    {\tiny  \citet{paper114}}  & a  &    & b  & c  & b  & b  &  1 & b  &    & b  & b  & abdh  \\
    {\tiny  \citet{paper116}}  & b  &   ab & a  & b  & b  & b  &  1 & b  & b  & a  & a  &   ab  \\
    {\tiny  \citet{paper117}}  & a  &  a & b  & c  & a  & b  &  2 & b  & b  & a  & a  & a  \\
    {\tiny  \citet{paper118}}  & a  &  a & b  & c  & b  & b  &  4 & b  & b  & c  & a  &  adg  \\
    {\tiny  \citet{paper119}}  & a  &  i & a  & c  & a  & b  &  2 & b  & a  & a  & a  &     abdei  \\
    {\tiny  \citet{paper120}}  & a  &  j & a  & c  & a  & b  &  2 & b  & a  & b  & a  & aehj  \\
    {\tiny  \citet{paper121}}  & a  &  k & b  & c  & b  & b  &  1 & a  & b  & c  & b  &    \\
    {\tiny  \citet{paper122}}  & c  &   ed & b  & c  & a  & b  &  1 & b  & b  & c  & b  &   de  \\
    {\tiny  \citet{paper123}}  & d  & Ka & a  &  c & b  & b  &  8 &    & d  & b  & a  &    \\
    {\tiny  \citet{paper124}}  & a  &   ed & b  &  c & a  & b  &  1 & b  & b  & a  & b  &   de  \\
    {\tiny  \citet{paper125}}  & d  &    &    &    & a  & b  &    &    &    &    &    &   eh  \\
    {\tiny  \citet{paper126}}  & a  &  l & a  &  c & a  & b  &  5 & a  & a  & b  & a  & abdi  \\
    {\tiny  \citet{paper127}}  & a  &  k & a  & b  & a  & b  &  2 & b  & a  & b  & a  &  ade  \\
    {\tiny  \citet{paper128}}  & c  &    &    &    &    &    &    &    &    &    &    &    \\
    {\tiny  \citet{paper129}}  & a  &  k & c  &  c & b  & b  &  2 & b  & c  & c  & a  & a  \\
    {\tiny  \citet{paper130}}  & g  &  d & b  & a  & a  & b  &  5 & a  & c  & b  & b  &   de  \\
    {\tiny  \citet{paper131}}  & b  &    & c  &  c & a  & b  &  5 & b  & a  & a  & a  &  ade  \\
    {\tiny  \citet{paper133}}  & e  &    & b  &  c & a  & b  &  1 & b  & b  & c  & b  & e  \\
    {\tiny  \citet{paper134}}  & a  &  j & b  & a  & b  & b  &  1 & b  &    & b  & b  & j  \\
    {\tiny  \citet{paper136}}  & a  &  l & b  &  c & a  & a  &  1 & b  & a  & c  & b  &   ej  \\
    {\tiny  \citet{paper137}}  & d  &    & a  &  c & a  & b  &  2 & b  & a  & b  & a  &  ade  \\
    {\tiny  \citet{paper138}}  & a  &  b & a  &  c & a  & b  &  2 & b  & a  & b  & a  &  abg  \\
    {\tiny  \citet{paper139}}  & a  &  j & a  &  c & b  & b  &  3 & b  & a  & b  & a  & j  \\
    {\tiny  \citet{paper140}}  & b  &  e & b  &  c & b  & b  &  1 & a  & b  & b  & b  & e  \\
    {\tiny  \citet{paper141}}  & b  &  k & a  & a  & b  & b  &  6 & b  & a  & c  & a  & bdei  \\
    {\tiny  \citet{paper142}}  & a  &  b & a  &  c & a  & b  &  2 & b  & a  & b  & a  &    abdegh  \\
    {\tiny  \citet{paper143}}  & a  &  k & a  &  c & a  & b  &  6 & b  & a  & c  & b  & abde  \\
    {\tiny  \citet{paper144}}  & a  &  l & c  &  c & b  & b  &  3 & a  &    & d  & b  & e  \\
    {\tiny  \citet{paper146}}  & a  &  j & b  &  c & a  & b  &  1 & b  & a  & c  & b  & aehj  \\
    {\tiny  \citet{paper147}}  & b  &   ba &    &    &    &    &    &    &    &    &    &   ab  \\
    {\tiny  \citet{paper148}}  & a  &  j & b  &  c & b  & b  &  2 & a  & e  & a  & b  & e  \\
    {\tiny  \citet{paper149}}  & a  &  d & b  &  c & b  & b  &  1 & b  & b  & b  & b  &     abdeh  \\
    {\tiny  \citet{paper150}}  & a  &  g & b  & a  & b  & b  &  5 & a  & b  & b  & b  &   dg  \\
    {\tiny  \citet{papern1}}  &  a &  k & a  &  c &  b &  b &  1 &  b &  b &  c &  a &  a \\
    {\tiny  \citet{papern2}}  &  b &  k & a  &  c &  b &  b &  1 &  b &  a &  b &  a & hejbda \\
    {\tiny  \citet{papern3}}  &  c &  l &    &    &    &    &    &    &    &    &  b &   jb \\
    {\tiny  \citet{papern4}}  &  c &  d & b  &  c &  b &  b &  3 &  b &  c &  b &  b &  d \\
    {\tiny  \citet{papern5}}  &  a &  a & a  &  c &  b &  b &  1 &  b &  c &  b &  a &    \\
    {\tiny  \citet{papern6}}  &  a &  b & c  &  c &  b &  b &  2 &  b &  a &  c &  a & ebj \\
    {\tiny  \citet{papern7}}  &  a &  k & b  &  c &  a &  b &  1 &  a &  b &  b &  b &    daib \\
    {\tiny  \citet{papern8}}  &  e &    &    &    &    &    &    &    &    &    &  b &    \\
    {\tiny  \citet{papern9}}  &  d &  i & b  &  b &  a &  b &  1 &  b &  b &    &  b &    \\
    {\tiny  \citet{papern10}}  &  c &    & b  &  c &  b &  b &  1 &  b &  a &  a &  b &   ea \\
    {\tiny  \citet{papern11}}  &  a &  l & a  &  c &  b &  b &  1 &  b &  a &  b &  a &   ba \\
    {\tiny  \citet{papern12}}  &  a &  l & b  &  c &  b &  b &  4 &  b &  a &  b &  a &    bade \\
    {\tiny  \citet{papern13}}  &  a &  j & b  &  c &  b &  b &  1 &  b &  b &    &  b &    \\
    {\tiny  \citet{papern15}}  &  c &    & b  &    &    &    &  1 &  b &  b &  b &  b &  e \\
    {\tiny  \citet{papern16}}  &  c &  e & a  &  b &  b &  b &    &  a &  b &  b &  b &  e \\
    {\tiny  \citet{papern17}}  &  a &  j & c  &  c &  a &  b &  3 &  b &  b &  b &  a &  f \\
    {\tiny  \citet{papern18}}  &  c &  e &    &    &    &    &    &    &    &    &  b &  e \\
    {\tiny  \citet{papern19}}  &  a &  l & b  &    &  b &  a &  1 &    &  a &    &  b &    \\
    {\tiny  \citet{papern21}}  &  a &  j & b  &  b &  b &  b &  1 &    &  b &    &  b & eda \\
    {\tiny  \citet{papern22}}  &  f &    &    &    &    &    &    &    &    &  b &  b &   ea \\
    {\tiny  \citet{papern24}}  &  g &    &    &    &    &  a &    &  b &    &    &  b &    \\
    {\tiny  \citet{papern25}}  &  a &  l & c  &  c &  b &  b &  3 &  a &  a &  a &  a &   bf \\
    {\tiny  \citet{papern26}}  &  a &  k & a  &  c &  b &  b &  3 &    &  b &  a &  b &    \\
    {\tiny  \citet{papern27}}  &  a &  j & b  &    &  b &  b &  1 &    &  a &    &  b &   eb \\
    {\tiny  \citet{papern29}}  &  a &  c &    &    &    &    &    &    &    &  b &  b &    \\
    {\tiny  \citet{papern31}}  &  d &    & a  &  c &  a &  b &  4 &  b &  a &  b &  a &  ehgdi \\
    {\tiny  \citet{papern32}}  &  b &  k & a  &  c &  b &  b &  3 &  b &  a &  b &  a & bad \\
    {\tiny  \citet{papern33}}  &  b &    &    &    &    &    &    &    &    &    &  b &   id \\
    {\tiny  \citet{papern35}}  &  a &  k & a  &  c &  b &  b &  5 &  b &  a &  b &  a &    aejd \\
    {\tiny  \citet{papern37}}  &  a &  a & b  &  c &  b &  b &  1 &  b &  b &  b &  b &  e \\
    {\tiny  \citet{papern38}}  &  a &  l & b  &  b &  b &  b &  1 &  b &  b &    &  b &    \\
    {\tiny  \citet{papern40}}  &  a &  k & c  &  c &  b &  a &  2 &  a &  a &  b &  b &  b \\
    {\tiny  \citet{papern42}}  &  e &    &    &    &    &    &    &    &    &    &  b &    \\
    {\tiny  \citet{papern43}}  &  a &  a & c  &  c &  a &  b &  1 &  b &  b &  d &  a &    \\
    {\tiny  \citet{papern45}}  &  c &    &    &    &    &    &    &    &    &  d &  b &    \\
    {\tiny  \citet{papern46}}  &  a &  l & a  &  c &  b &  b &  2 &    &  a &  b &  a &  hajbe \\
    {\tiny  \citet{papern47}}  &  d &    & c  &  c &  a &  b &  3 &    &    &  b &  a &    eiad \\
    {\tiny  \citet{papern49}}  &  a &  c & a  &  c &  a &  b &  1 &  a &  a &  a &  a &    \\
    {\tiny  \citet{papern50}}  &  a &  a & b  &  c &  b &  b &  1 &  b &  a &  b &  b &  hjabd \\
    {\tiny  \citet{papern51}}  &  a &  e & b  &  c &  a &  b &  1 &  b &  b &  b &  b &  e \\
    {\tiny  \citet{papern52}}  &  c &    &    &    &    &    &    &    &    &    &    &    \\
    {\tiny  \citet{papern53}}  &  a &  c &    &    &    &    &    &    &    &    &    &    \\
    {\tiny  \citet{papern54}}  &  a &  e & b  &  c &  a &  b &  1 &  b &  b &  d &  b & ejd \\
    {\tiny  \citet{papern55}}  &  a &  l & a  &  c &  b &  b &  9 &  b &  c &  d &  a &  ihejb \\
    {\tiny  \citet{papern56}}  &  a &  i &    &    &  b &  b &    &  b &    &    &  b &    \\
    {\tiny  \citet{papern57}}  &  a &  b & a  &  c &  a &  b & 10 &  a &  b &  b &  a &  b \\
    {\tiny  \citet{papern58}}  &  a &  b & a  &  c &  a &  b &  2 &  b &  a &  b &  a &  b \\
    {\tiny  \citet{papern59}}  &  c &  e &    &    &    &    &    &    &  b &  d &  b &    \\
    {\tiny  \citet{papern60}}  &  a &  b & b  &  c &  b &  b &  1 &  b &  b &  b &  b &  b \\
    {\tiny  \citet{papern61}}  &  a &  j & a  &  c &  a &  a &  9 &    &  a &  b &  a &    \\
    {\tiny  \citet{papern62}}  &  e &    &    &    &    &    &    &    &    &    &    &    \\
    {\tiny  \citet{papern63}}  &  a &  j & a  &  c &  a &  b &  1 &  b &  a &  c &  a & bea \\
    {\tiny  \citet{papern64}}  &  a &  a & b  &  c &  a &  b &  1 &  b &  a &  b &  b & eha \\
    {\tiny  \citet{papern65}}  &  c &    & b  &  c &  a &  b &  1 &  b &  b &    &    &    \\
    
\label{tab:papers_all}%

\end{longtable}%
}


\begin{thebibliography}{215}
\expandafter\ifx\csname natexlab\endcsname\relax\def\natexlab#1{#1}\fi
\expandafter\ifx\csname url\endcsname\relax
  \def\url#1{\texttt{#1}}\fi
\expandafter\ifx\csname urlprefix\endcsname\relax\def\urlprefix{URL }\fi

\bibitem[{Abdou(2009)}]{paper81}
Abdou, H., 2009. Genetic programming for credit scoring: The case of egyptian
  public sector banks. Expert Systems with Applications 36~(9), 11402--11417.

\bibitem[{Abdou et~al.(2008)Abdou, Pointon, and El-Masry}]{paper64}
Abdou, H., Pointon, J., El-Masry, A., 2008. Neural nets versus conventional
  techniques in credit scoring in egyptian banking. Expert Systems with
  Applications 35~(3), 1275--1292.

\bibitem[{Abdoun and Abouchabaka(2011)}]{abdoun2012comparative}
Abdoun, O., Abouchabaka, J., 2011. A comparative study of adaptive crossover
  operators for genetic algorithms to resolve the traveling salesman problem.
  International Journal of Computer Applications 31~(11), 49--57.

\bibitem[{Abell{\`a}n and Mantas(2014)}]{papern26}
Abell{\`a}n, J.n, J., Mantas, C., 2014. Improving experimental studies about
  ensembles of classifiers for bankruptcy prediction and credit scoring. Expert
  Systems with Applications 41~(8), 3825--3830.

\bibitem[{Adams et~al.(2001)Adams, Hand, and Till}]{paper10}
Adams, N., Hand, D., Till, R., 2001. Mining for classes and patterns in
  behavioural data. Journal of the Operational Research Society 52~(9),
  1017--1024.

\bibitem[{Akkoc(2012)}]{paper146}
Akkoc, S., 2012. An empirical comparison of conventional techniques, neural
  networks and the three stage hybrid adaptive neuro fuzzy inference system
  (anfis) model for credit scoring analysis: The case of turkish credit card
  data. European Journal of Operational Research 222~(1), 168--178.

\bibitem[{Antonakis and Sfakianakis(2009)}]{paper75}
Antonakis, A., Sfakianakis, M., 2009. Assessing naivee bayes as a method for
  screening credit applicants. Journal of Applied Statistics 36~(5), 537--545.

\bibitem[{Aryuni and Madyatmadja(2015)}]{papern9}
Aryuni, M., Madyatmadja, E., 2015. Feature selection in credit scoring model
  for credit card applicants in xyz bank: A comparative study. International
  Journal of Multimedia and Ubiquitous Engineering 10~(5), 17--24.

\bibitem[{Bache and Lichman(2013)}]{uci:2013}
Bache, K., Lichman, M., 2013. {UCI} machine learning repository.
\newline\urlprefix\url{http://archive.ics.uci.edu/ml}

\bibitem[{Baesens et~al.(2005)Baesens, Van~Gestel, Stepanova, Van Den~Poel, and
  Vanthienen}]{paper33}
Baesens, B., Van~Gestel, T., Stepanova, M., Van Den~Poel, D., Vanthienen, J.,
  2005. Neural network survival analysis for personal loan data. Journal of the
  Operational Research Society 56~(9), 1089--1098.

\bibitem[{Baesens et~al.(2003)Baesens, Van~Gestel, Viaene, Stepanova, Suykens,
  and Vanthienen}]{paper21}
Baesens, B., Van~Gestel, T., Viaene, S., Stepanova, M., Suykens, J.,
  Vanthienen, J., 2003. Benchmarking state-of-the-art classification algorithms
  for credit scoring. Journal of the Operational Research Society 54~(6),
  627--635.

\bibitem[{Bahnsen et~al.(2015)Bahnsen, Aouada, and Ottersten}]{papern4}
Bahnsen, A., Aouada, D., Ottersten, B., 2015. Example-dependent cost-sensitive
  decision trees. Expert Systems with Applications 42~(19), 6609--6619.

\bibitem[{Banasik et~al.(1999)Banasik, Crook, and Thomas}]{paper6}
Banasik, J., Crook, J., Thomas, L., 1999. Not if but when will borrowers
  default. Journal of the Operational Research Society 50~(12), 1185--1190.

\bibitem[{Banasik et~al.(2003)Banasik, Crook, and Thomas}]{paper22}
Banasik, J., Crook, J., Thomas, L., 2003. Sample selection bias in credit
  scoring models. Journal of the Operational Research Society 54~(8), 822--832.

\bibitem[{Bardos(1998)}]{paper4}
Bardos, M., 1998. Detecting the risk of company failure at the banque de
  france. Journal of Banking and Finance 22~(10-11), 1405--1419.

\bibitem[{Baxter et~al.(2007)Baxter, Gawler, and Ang}]{paper53}
Baxter, R., Gawler, M., Ang, R., 2007. Predictive model of insolvency risk for
  australian corporations. Conferences in Research and Practice in Information
  Technology Series 70, 21--28.

\bibitem[{Bekhet and Eletter(2014)}]{papern37}
Bekhet, H., Eletter, S., 2014. Credit risk assessment model for jordanian
  commercial banks: Neural scoring approach. Review of Development Finance
  4~(1), 20--28.

\bibitem[{Ben-David and Frank(2009)}]{paper73}
Ben-David, A., Frank, E., 2009. Accuracy of machine learning models versus hand
  craft. Expert Systems with Applications 36~(3 PART 1), 5264--5271.

\bibitem[{Berger et~al.(2005)Berger, Frame, and Miller}]{paper26}
Berger, A., Frame, W., Miller, N., 2005. Credit scoring and the availability,
  price, and risk of small business credit. Journal of Money, Credit and
  Banking 37~(2), 191--222.

\bibitem[{Berkson(1944)}]{berkson1944application}
Berkson, J., 1944. Application of the logistic function to bio-assay. Journal
  of the American Statistical Association 39~(227), 357--365.

\bibitem[{Bijak and Thomas(2012)}]{paper130}
Bijak, K., Thomas, L., 2012. Does segmentation always improve model performance
  in credit scoring? Expert Systems with Applications 39~(3), 2433--2442.

\bibitem[{Bishop(1995)}]{bishop1995neural}
Bishop, C.~M., 1995. Neural networks for pattern recognition. Oxford university
  press.

\bibitem[{Blanco et~al.(2013)Blanco, Pino-Mej{\'i}as, Lara, and
  Rayo}]{papern64}
Blanco, A., Pino-Mej{\'i}as, R., Lara, J., Rayo, S., 2013. Credit scoring
  models for the microfinance industry using neural networks: Evidence from
  peru. Expert Systems with Applications 40~(1), 356--364.

\bibitem[{Bravo and Maldonado(2015)}]{papern18}
Bravo, C., Maldonado, S., 2015. Fieller stability measure: A novel
  model-dependent backtesting approach. Journal of the Operational Research
  Society 66~(11), 1895--1905.

\bibitem[{Bravo et~al.(2013)Bravo, Maldonado, and Weber}]{papern52}
Bravo, C., Maldonado, S., Weber, R., 2013. Granting and managing loans for
  micro-entrepreneurs: New developments and practical experiences. European
  Journal of Operational Research 227~(2), 358--366.

\bibitem[{Bravo et~al.(2015)Bravo, Thomas, and Weber}]{papern10}
Bravo, C., Thomas, L., Weber, R., 2015. Improving credit scoring by
  differentiating defaulter behaviour. Journal of the Operational Research
  Society 66~(5), 771--781.

\bibitem[{Breiman(1996)}]{breiman1996}
Breiman, L., 1996. Bagging predictors. Machine learning 24~(2), 123--140.

\bibitem[{Breiman et~al.(1984)Breiman, Friedman, Olshen, and
  Stone}]{breiman1984classification}
Breiman, L., Friedman, J.~H., Olshen, R.~A., Stone, C.~J., 1984. Classification
  and regression trees. wadsworth \& brooks. Monterey, CA.

\bibitem[{Brown and Mues(2012)}]{paper131}
Brown, I., Mues, C., 2012. An experimental comparison of classification
  algorithms for imbalanced credit scoring data sets. Expert Systems with
  Applications 39~(3), 3446--3453.

\bibitem[{Burton(2012)}]{paper128}
Burton, D., 2012. Credit scoring, risk, and consumer lendingscapes in emerging
  markets. Environment and Planning A 44~(1), 111--124.

\bibitem[{Capotorti and Barbanera(2012)}]{paper134}
Capotorti, A., Barbanera, E., 2012. Credit scoring analysis using a fuzzy
  probabilistic rough set model. Computational Statistics and Data Analysis
  56~(4), 981--994.

\bibitem[{Chang and Yeh(2012)}]{paper126}
Chang, S.-Y., Yeh, T.-Y., 2012. An artificial immune classifier for credit
  scoring analysis. Applied Soft Computing Journal 12~(2), 611--618.

\bibitem[{Chen and Li(2010)}]{paper102}
Chen, F.-L., Li, F.-C., 2010. Combination of feature selection approaches with
  svm in credit scoring. Expert Systems with Applications 37~(7), 4902--4909.

\bibitem[{Chen and Huang(2003)}]{paper20}
Chen, M.-C., Huang, S.-H., 2003. Credit scoring and rejected instances
  reassigning through evolutionary computation techniques. Expert Systems with
  Applications 24~(4), 433--441.

\bibitem[{Chen et~al.(2009)Chen, Ma, and Ma}]{paper74}
Chen, W., Ma, C., Ma, L., 2009. Mining the customer credit using hybrid support
  vector machine technique. Expert Systems with Applications 36~(4),
  7611--7616.

\bibitem[{Chrzanowska et~al.(2009)Chrzanowska, Alfaro, and Witkowska}]{paper72}
Chrzanowska, M., Alfaro, E., Witkowska, D., 2009. The individual borrowers
  recognition: Single and ensemble trees. Expert Systems with Applications
  36~(3 PART 2), 6409--6414.

\bibitem[{Chuang and Huang(2011)}]{paper113}
Chuang, C.-L., Huang, S.-T., 2011. A hybrid neural network approach for credit
  scoring. Expert Systems 28~(2), 185--196.

\bibitem[{Chuang and Lin(2009)}]{paper67}
Chuang, C.-L., Lin, R.-H., 2009. Constructing a reassigning credit scoring
  model. Expert Systems with Applications 36~(2 PART 1), 1685--1694.

\bibitem[{Cubiles-De-La-Vega et~al.(2013)Cubiles-De-La-Vega, Blanco-Oliver,
  Pino-Mej{\'i}as, and Lara-Rubio}]{papern49}
Cubiles-De-La-Vega, M.-D., Blanco-Oliver, A., Pino-Mej{\'i}as, R., Lara-Rubio,
  J., 2013. Improving the management of microfinance institutions by using
  credit scoring models based on statistical learning techniques. Expert
  Systems with Applications 40~(17), 6910--6917.

\bibitem[{Deng et~al.(2015)Deng, Huye, He, Li, and Li}]{papern11}
Deng, Z., Huye, B., He, P., Li, Y., Li, P., 2015. An artificial immune network
  classification algorithm for credit scoring. Journal of Information and
  Computational Science 12~(11), 4263--4270.

\bibitem[{DeYoung et~al.(2011)DeYoung, Frame, Glennon, and Nigro}]{paper110}
DeYoung, R., Frame, W., Glennon, D., Nigro, P., 2011. The information
  revolution and small business lending: The missing evidence. Journal of
  Financial Services Research 39~(41306), 19--33.

\bibitem[{Dong et~al.(2011)Dong, Hao, and Yu}]{paper114}
Dong, Y., Hao, X., Yu, C., 2011. Comparison of statistical and artificial
  intelligence methodologies in small-businesses' credit assessment based on
  daily transaction data. ICIC Express Letters 5~(5), 1725--1730.

\bibitem[{Dryver and Sukkasem(2009)}]{paper66}
Dryver, A., Sukkasem, J., 2009. Validating risk models with a focus on credit
  scoring models. Journal of Statistical Computation and Simulation 79~(2),
  181--193.

\bibitem[{Durand(1941)}]{Durand41}
Durand, D., 1941. Risk elements in consumer instalment financing. In: National
  Bureau of Economics. New York.

\bibitem[{Efromovich(2010)}]{paper93}
Efromovich, S., 2010. Oracle inequality for conditional density estimation and
  an actuarial example. Annals of the Institute of Statistical Mathematics
  62~(2), 249--275.

\bibitem[{Einav et~al.(2013)Einav, Jenkins, and Levin}]{papern53}
Einav, L., Jenkins, M., Levin, J., 2013. The impact of credit scoring on
  consumer lending. RAND Journal of Economics 44~(2), 249--274.

\bibitem[{Falangis and Glen(2010)}]{paper98}
Falangis, K., Glen, J., 2010. Heuristics for feature selection in mathematical
  programming discriminant analysis models. Journal of the Operational Research
  Society 61~(5), 804--812.

\bibitem[{Feng et~al.(2010)Feng, Yao, and Jin}]{paper104}
Feng, L., Yao, Y., Jin, B., 2010. Research on credit scoring model with svm for
  network management. Journal of Computational Information Systems 6~(11),
  3567--3574.

\bibitem[{Ferreira et~al.(2015)Ferreira, Louzada, and Diniz}]{papern16}
Ferreira, P., Louzada, F., Diniz, C., 2015. Credit scoring modeling with
  state-dependent sample selection: A comparison study with the usual logistic
  modeling. Pesquisa Operacional 35~(1), 39--56.

\bibitem[{Figini and Uberti(2010)}]{paper88}
Figini, S., Uberti, P., 2010. Model assessment for predictive classification
  models. Communications in Statistics - Theory and Methods 39~(18),
  3238--3244.

\bibitem[{Finlay(2008)}]{paper62}
Finlay, S., 2008. Towards profitability: A utility approach to the credit
  scoring problem. Journal of the Operational Research Society 59~(7),
  921--931.

\bibitem[{Finlay(2009)}]{paper77}
Finlay, S., 2009. Are we modelling the right thing? the impact of incorrect
  problem specification in credit scoring. Expert Systems with Applications
  36~(5), 9065--9071.

\bibitem[{Finlay(2010)}]{paper97}
Finlay, S., 2010. Credit scoring for profitability objectives. European Journal
  of Operational Research 202~(2), 528--537.

\bibitem[{Finlay(2011)}]{paper112}
Finlay, S., 2011. Multiple classifier architectures and their application to
  credit risk assessment. European Journal of Operational Research 210~(2),
  368--378.

\bibitem[{Fisher(1986)}]{FISHER}
Fisher, R.~A., 1986. The use of multiple measurements in taxonomic problems.
  Annals of Eugenics 7, 179--188.

\bibitem[{Florez-Lopez and Ramon-Jeronimo(2015)}]{papern2}
Florez-Lopez, R., Ramon-Jeronimo, J., 2015. Enhancing accuracy and
  interpretability of ensemble strategies in credit risk assessment. a
  correlated-adjusted decision forest proposal. Expert Systems with
  Applications 42~(13), 5737--5753.

\bibitem[{Friedman et~al.(1997)Friedman, Geiger, and Goldszmidt}]{FRIEDMAN}
Friedman, N., Geiger, D., Goldszmidt, M., 1997. Bayesian network classifiers.
  Machine Learning 29(2-3), 131--163.

\bibitem[{Garcia et~al.(2014)Garcia, Marques, and Sanchez}]{papern42}
Garcia, V., Marques, A., Sanchez, J., 2014. An insight into the experimental
  design for credit risk and corporate bankruptcy prediction systems. Journal
  of Intelligent Information Systems 44~(1), 159--189.

\bibitem[{Gemela(2001)}]{paper13}
Gemela, J., 2001. Financial analysis using bayesian networks. Applied
  Stochastic Models in Business and Industry 17~(1), 57--67.

\bibitem[{Gestel et~al.(2006)Gestel, Baesens, Suykens, Van~den Poel, Baestaens,
  and Willekens}]{paper43}
Gestel, T., Baesens, B., Suykens, J., Van~den Poel, D., Baestaens, D.-E.,
  Willekens, M., 2006. Bayesian kernel based classification for financial
  distress detection. European Journal of Operational Research 172~(3),
  979--1003.

\bibitem[{Giudici(2001)}]{paper8}
Giudici, P., 2001. Bayesian data mining, with application to benchmarking and
  credit scoring. Applied Stochastic Models in Business and Industry 17~(1),
  69--81.

\bibitem[{Gzyl et~al.(2015)Gzyl, Ter~Horst, and Molina}]{papern3}
Gzyl, H., Ter~Horst, E., Molina, G., 2015. Application of the method of maximum
  entropy in the mean to classification problems. Physica A: Statistical
  Mechanics and its Applications 437, 101--108.

\bibitem[{Hachicha and Ghorbel(2012)}]{hachicha2012survey}
Hachicha, W., Ghorbel, A., 2012. A survey of control-chart pattern-recognition
  literature (1991--2010) based on a new conceptual classification scheme.
  Computers \& Industrial Engineering 63~(1), 204--222.

\bibitem[{Han et~al.(2006)Han, Kamber, and Pei}]{han2006data}
Han, J., Kamber, M., Pei, J., 2006. Data mining: concepts and techniques.
  Morgan kaufmann.

\bibitem[{Hand(2001{\natexlab{a}})}]{paper11}
Hand, D., 2001{\natexlab{a}}. Modelling consumer credit risk. IMA Journal
  Management Mathematics 12~(2), 139--155.

\bibitem[{Hand(2001{\natexlab{b}})}]{Hand2001a}
Hand, D., 2001{\natexlab{b}}. Modelling consumer credit risk. IMA Journal
  Management Mathematics 12~(2), 139--155.

\bibitem[{Hand(2005{\natexlab{a}})}]{paper34}
Hand, D., 2005{\natexlab{a}}. Good practice in retail credit scorecard
  assessment. Journal of the Operational Research Society 56~(9), 1109--1117.

\bibitem[{Hand(2005{\natexlab{b}})}]{paper25}
Hand, D., 2005{\natexlab{b}}. Supervised classification and tunnel vision.
  Applied Stochastic Models in Business and Industry 21~(2), 97--109.

\bibitem[{Hand(2006)}]{Hand2006}
Hand, D., 2006. Classifier technology and the illusion of progress. Statistical
  Science 21~(9), 1--14.

\bibitem[{Hand and Henley(1997)}]{paper3}
Hand, D., Henley, W., 1997. Statistical classification methods in consumer
  credit scoring: A review. Journal of the Royal Statistical Society. Series A:
  Statistics in Society 160~(3), 523--541.

\bibitem[{Hand and Kelly(2002)}]{paper18}
Hand, D., Kelly, M., 2002. Superscorecards. IMA Journal Management Mathematics
  13~(4), 273--281.

\bibitem[{Hardle et~al.(1998)Hardle, Mammen, and Muller}]{paper5}
Hardle, W., Mammen, E., Muller, M., 1998. Testing parametric versus
  semiparametric modeling in generalized linear models. Journal of the American
  Statistical Association 93~(444), 1461--1474.

\bibitem[{Harris(2015)}]{papern6}
Harris, T., 2015. Credit scoring using the clustered support vector machine.
  Expert Systems with Applications 42~(2), 741--750.

\bibitem[{He et~al.(2010)He, Zhang, Shi, and Huang}]{paper100}
He, J., Zhang, Y., Shi, Y., Huang, G., 2010. Domain-driven classification based
  on multiple criteria and multiple constraint-level programming for
  intelligent credit scoring. IEEE Transactions on Knowledge and Data
  Engineering 22~(6), 826--838.

\bibitem[{Hens and Tiwari(2012)}]{paper138}
Hens, A., Tiwari, M., 2012. Computational time reduction for credit scoring: An
  integrated approach based on support vector machine and stratified sampling
  method. Expert Systems with Applications 39~(8), 6774--6781.

\bibitem[{Hofer(2015)}]{papern15}
Hofer, V., 2015. Adapting a classification rule to local and global shift when
  only unlabelled data are available. European Journal of Operational Research
  243~(1), 177--189.

\bibitem[{Hofer and Krempl(2013)}]{papern65}
Hofer, V., Krempl, G., 2013. Drift mining in data: A framework for addressing
  drift in classification. Computational Statistics and Data Analysis 57~(1),
  377--391.

\bibitem[{Hoffmann et~al.(2002)Hoffmann, Baesens, Martens, Put, and
  Vanthienen}]{paper19}
Hoffmann, F., Baesens, B., Martens, J., Put, F., Vanthienen, J., 2002.
  Comparing a genetic fuzzy and a neurofuzzy classifier for credit scoring.
  International Journal of Intelligent Systems 17~(11), 1067--1083.

\bibitem[{Hoffmann et~al.(2007)Hoffmann, Baesens, Mues, Van, and
  Vanthienen}]{paper48}
Hoffmann, F., Baesens, B., Mues, C., Van, Gestel, T., Vanthienen, J., 2007.
  Inferring descriptive and approximate fuzzy rules for credit scoring using
  evolutionary algorithms. European Journal of Operational Research 177~(1),
  540--555.

\bibitem[{Hsieh(2005)}]{paper30}
Hsieh, N.-C., 2005. Hybrid mining approach in the design of credit scoring
  models. Expert Systems with Applications 28~(4), 655--665.

\bibitem[{Hsieh and Hung(2010)}]{paper89}
Hsieh, N.-C., Hung, L.-P., 2010. A data driven ensemble classifier for credit
  scoring analysis. Expert Systems with Applications 37~(1), 534--545.

\bibitem[{Hu and Ansell(2007)}]{paper55}
Hu, Y.-C., Ansell, J., 2007. Measuring retail company performance using credit
  scoring techniques. European Journal of Operational Research 183~(3),
  1595--1606.

\bibitem[{Hu and Ansell(2009)}]{paper83}
Hu, Y.-C., Ansell, J., 2009. Retail default prediction by using sequential
  minimal optimization technique. Journal of Forecasting 28~(8), 651--666.

\bibitem[{Huang et~al.(2007)Huang, Chen, and Wang}]{paper52}
Huang, C.-L., Chen, M.-C., Wang, C.-J., 2007. Credit scoring with a data mining
  approach based on support vector machines. Expert Systems with Applications
  33~(4), 847--856.

\bibitem[{Huang et~al.(2006{\natexlab{a}})Huang, Tzeng, and Ong}]{paper39}
Huang, J.-J., Tzeng, G.-H., Ong, C.-S., 2006{\natexlab{a}}. Two-stage genetic
  programming {(2SGP)} for the credit scoring model. Applied Mathematics and
  Computation 174~(2), 1039--1053.

\bibitem[{Huang et~al.(2006{\natexlab{b}})Huang, Hung, and Jiau}]{paper44}
Huang, Y.-M., Hung, C.-M., Jiau, H., 2006{\natexlab{b}}. Evaluation of neural
  networks and data mining methods on a credit assessment task for class
  imbalance problem. Nonlinear Analysis: Real World Applications 7~(4),
  720--747.

\bibitem[{Huysmans et~al.(2006)Huysmans, Baesens, Vanthienen, and
  Van~Gestel}]{paper40}
Huysmans, J., Baesens, B., Vanthienen, J., Van~Gestel, T., 2006. Failure
  prediction with self organizing maps. Expert Systems with Applications
  30~(3), 479--487.

\bibitem[{John et~al.(1996)John, Miller, and Kerber}]{paper1}
John, G., Miller, P., Kerber, R., 1996. Stock selection using rule induction.
  IEEE Expert-Intelligent Systems and their Applications 11~(5), 52--58.

\bibitem[{Jung and Thomas(2008)}]{paper60}
Jung, K., Thomas, L., 2008. A note on coarse classifying in acceptance
  scorecards. Journal of the Operational Research Society 59~(5), 714--718.

\bibitem[{Kao et~al.(2012)Kao, Chiu, and Chiu}]{paper149}
Kao, L.-J., Chiu, C.-C., Chiu, F.-Y., 2012. A bayesian latent variable model
  with classification and regression tree approach for behavior and credit
  scoring. Knowledge-Based Systems 36, 245--252.

\bibitem[{Karlis and Rahmouni(2007)}]{paper49}
Karlis, D., Rahmouni, M., 2007. Analysis of defaulters' behaviour using the
  poisson-mixture approach. IMA Journal Management Mathematics 18~(3),
  297--311.

\bibitem[{Kennedy et~al.(2013{\natexlab{a}})Kennedy, Mac~Namee, Delany,
  O'Sullivan, and Watson}]{papern59}
Kennedy, K., Mac~Namee, B., Delany, S., O'Sullivan, M., Watson, N.,
  2013{\natexlab{a}}. A window of opportunity: Assessing behavioural scoring.
  Expert Systems with Applications 40~(4), 1372--1380.

\bibitem[{Kennedy et~al.(2013{\natexlab{b}})Kennedy, Namee, and
  Delany}]{papern55}
Kennedy, K., Namee, B., Delany, S., 2013{\natexlab{b}}. Using semi-supervised
  classifiers for credit scoring. Journal of the Operational Research Society
  64~(4), 513--529.

\bibitem[{Khashei et~al.(2013)Khashei, Rezvan, Hamadani, and Bijari}]{papern50}
Khashei, M., Rezvan, M., Hamadani, A., Bijari, M., 2013. A bi-level
  neural-based fuzzy classification approach for credit scoring problems.
  Complexity 18~(6), 46--57.

\bibitem[{Kocenda and Vojtek(2011)}]{paper124}
Kocenda, E., Vojtek, M., 2011. Default predictors in retail credit scoring:
  Evidence from czech banking data. Emerging Markets Finance and Trade 47~(6),
  80--98.

\bibitem[{Kolbe and Brunette(1991)}]{kolbe1991}
Kolbe, R., Brunette, M., 1991. Content analysis research: An examination of
  applications with directives for improving research, reliability and
  objectivity. Journal of Consumer Research 18~(2), 243--250.

\bibitem[{Koutanaei et~al.(2015)Koutanaei, Sajedi, and Khanbabaei}]{papern7}
Koutanaei, F., Sajedi, H., Khanbabaei, M., 2015. A hybrid data mining model of
  feature selection algorithms and ensemble learning classifiers for credit
  scoring. Journal of Retailing and Consumer Services 27, 11--23.

\bibitem[{Koza(1992)}]{koza1992genetic}
Koza, J.~R., 1992. Genetic programming: on the programming of computers by
  means of natural selection (complex adaptive systems).

\bibitem[{Kruppa et~al.(2013)Kruppa, Schwarz, Arminger, and Ziegler}]{papern54}
Kruppa, J., Schwarz, A., Arminger, G., Ziegler, A., 2013. Consumer credit risk:
  Individual probability estimates using machine learning. Expert Systems with
  Applications 40~(13), 5125--5131.

\bibitem[{Laha(2007)}]{paper51}
Laha, A., 2007. Building contextual classifiers by integrating fuzzy rule based
  classification technique and k-nn method for credit scoring. Advanced
  Engineering Informatics 21~(3), 281--291.

\bibitem[{Lahsasna et~al.(2010{\natexlab{a}})Lahsasna, Ainon, and
  Wah}]{paper96}
Lahsasna, A., Ainon, R., Wah, T., 2010{\natexlab{a}}. Credit scoring models
  using soft computing methods: A survey. International Arab Journal of
  Information Technology 7~(2), 115--123.

\bibitem[{Lahsasna et~al.(2010{\natexlab{b}})Lahsasna, Ainon, and
  Wah}]{paper85}
Lahsasna, A., Ainon, R., Wah, T., 2010{\natexlab{b}}. Enhancement of
  transparency and accuracy of credit scoring models through genetic fuzzy
  classifier. Maejo International Journal of Science and Technology 4~(1),
  136--158.

\bibitem[{Lan et~al.(2006)Lan, Janssens, Chen, and Wets}]{paper42}
Lan, Y., Janssens, D., Chen, G., Wets, G., 2006. Improving associative
  classification by incorporating novel interestingness measures. Expert
  Systems with Applications 31~(1), 184--192.

\bibitem[{Lee and Chen(2005)}]{paper29}
Lee, T.-S., Chen, I.-F., 2005. A two-stage hybrid credit scoring model using
  artificial neural networks and multivariate adaptive regression splines.
  Expert Systems with Applications 28~(4), 743--752.

\bibitem[{Lee et~al.(2006)Lee, Chiu, Chou, and Lu}]{paper38}
Lee, T.-S., Chiu, C.-C., Chou, Y.-C., Lu, C.-J., 2006. Mining the customer
  credit using classification and regression tree and multivariate adaptive
  regression splines. Computational Statistics and Data Analysis 50~(4),
  1113--1130.

\bibitem[{Lee et~al.(2002)Lee, Chiu, Lu, and Chen}]{paper17}
Lee, T.-S., Chiu, C.-C., Lu, C.-J., Chen, I.-F., 2002. Credit scoring using the
  hybrid neural discriminant technique. Expert Systems with Applications
  23~(3), 245--254.

\bibitem[{Lessmann et~al.(2015)Lessmann, Baesens, Seow, and Thomas}]{papern8}
Lessmann, S., Baesens, B., Seow, H.-V., Thomas, L., 2015. Benchmarking
  state-of-the-art classification algorithms for credit scoring: An update of
  research. European Journal of Operational Research 247~(1), 124--136.

\bibitem[{Li and Hand(2002)}]{paper15}
Li, H., Hand, D., 2002. Direct versus indirect credit scoring classifications.
  Journal of the Operational Research Society 53~(6), 647--654.

\bibitem[{Li et~al.(2006)Li, Shiue, and Huang}]{paper41}
Li, S.-T., Shiue, W., Huang, M.-H., 2006. The evaluation of consumer loans
  using support vector machines. Expert Systems with Applications 30~(4),
  772--782.

\bibitem[{Liang et~al.(2014)Liang, Tsai, and Wu}]{papern31}
Liang, D., Tsai, C.-F., Wu, H.-T., 2014. The effect of feature selection on
  financial distress prediction. Knowledge-Based Systems 73~(1), 289--297.

\bibitem[{Ling et~al.(2012)Ling, Cao, and Zhang}]{paper142}
Ling, Y., Cao, Q., Zhang, H., 2012. Credit scoring using multi-kernel support
  vector machine and chaos particle swarm optimization. International Journal
  of Computational Intelligence and Applications 11~(3),
  12500198:1--12500198:13.

\bibitem[{Lisboa et~al.(2009)Lisboa, Etchells, Jarman, Arsene, Aung, Eleuteri,
  Taktak, Ambrogi, Boracchi, and Biganzoli}]{paper78}
Lisboa, P., Etchells, T., Jarman, I., Arsene, C., Aung, M., Eleuteri, A.,
  Taktak, A., Ambrogi, F., Boracchi, P., Biganzoli, E., 2009. Partial logistic
  artificial neural network for competing risks regularized with automatic
  relevance determination. IEEE Transactions on Neural Networks 20~(9),
  1403--1416.

\bibitem[{Little and Rubin(2002)}]{little2002statistical}
Little, R.~J., Rubin, D.~B., 2002. Statistical analysis with missing data.

\bibitem[{Liu et~al.(2015)Liu, Hua, and Lim}]{papern19}
Liu, F., Hua, Z., Lim, A., 2015. Identifying future defaulters: A hierarchical
  bayesian method. European Journal of Operational Research 241~(1), 202--211.

\bibitem[{Liu et~al.(2010)Liu, Fu, and Lin}]{paper107}
Liu, X., Fu, H., Lin, W., 2010. A modified support vector machine model for
  credit scoring. International Journal of Computational Intelligence Systems
  3~(6), 797--803.

\bibitem[{Liu and Schumann(2005)}]{paper35}
Liu, Y., Schumann, M., 2005. Data mining feature selection for credit scoring
  models. Journal of the Operational Research Society 56~(9), 1099--1108.

\bibitem[{Louzada et~al.(2011)Louzada, Anacleto-Junior, Candolo, and
  Mazucheli}]{paper121}
Louzada, F., Anacleto-Junior, O., Candolo, C., Mazucheli, J., 2011.
  Poly-bagging predictors for classification modelling for credit scoring.
  Expert Systems with Applications 38~(10), 12717--12720.

\bibitem[{Louzada et~al.(2012{\natexlab{a}})Louzada, Cancho, Roman, and
  Leite}]{paper144}
Louzada, F., Cancho, V., Roman, M., Leite, J., 2012{\natexlab{a}}. A new
  long-term lifetime distribution induced by a latent complementary risk
  framework. Journal of Applied Statistics 39~(10), 2209--2222.

\bibitem[{Louzada et~al.(2012{\natexlab{b}})Louzada, Ferreira-Silva, and
  Diniz}]{paper140}
Louzada, F., Ferreira-Silva, P., Diniz, C., 2012{\natexlab{b}}. On the impact
  of disproportional samples in credit scoring models: An application to a
  brazilian bank data. Expert Systems with Applications 39~(9), 8071--8078.

\bibitem[{Lu et~al.(2013)Lu, Liyan, and Hongwei}]{papern63}
Lu, H., Liyan, H., Hongwei, Z., 2013. Credit scoring model hybridizing
  artificial intelligence with logistic regression. Journal of Networks 8~(1),
  253--261.

\bibitem[{Lucas(2001)}]{paper12}
Lucas, A., 2001. Statistical challenges in credit card issuing. Applied
  Stochastic Models in Business and Industry 17~(1), 69--81.

\bibitem[{Luo et~al.(2009)Luo, Cheng, and Hsieh}]{paper76}
Luo, S.-T., Cheng, B.-W., Hsieh, C.-H., 2009. Prediction model building with
  clustering-launched classification and support vector machines in credit
  scoring. Expert Systems with Applications 36~(4), 7562--7566.

\bibitem[{Madyatmadja and Aryuni(2014)}]{papern33}
Madyatmadja, E., Aryuni, M., 2014. Comparative study of data mining model for
  credit card application scoring in bank. Journal of Theoretical and Applied
  Information Technology 59~(2), 269--274.

\bibitem[{Majeske and Lauer(2013)}]{papern56}
Majeske, K., Lauer, T., 2013. The bank loan approval decision from multiple
  perspectives. Expert Systems with Applications 40~(5), 1591--1598.

\bibitem[{Marcano-Cedeno et~al.(2011)Marcano-Cedeno, Marin-De-La-Barcena,
  Jimenez-Trillo, Pinuela, and Andina}]{paper117}
Marcano-Cedeno, A., Marin-De-La-Barcena, A., Jimenez-Trillo, J., Pinuela, J.,
  Andina, D., 2011. Artificial metaplasticity neural network applied to credit
  scoring. International Journal of Neural Systems 21~(4), 311--317.

\bibitem[{Marques et~al.(2012{\natexlab{a}})Marques, Garcia, and
  Sanchez}]{paper141}
Marques, A., Garcia, V., Sanchez, J., 2012{\natexlab{a}}. Exploring the
  behaviour of base classifiers in credit scoring ensembles. Expert Systems
  with Applications 39~(11), 10244--10250.

\bibitem[{Marques et~al.(2012{\natexlab{b}})Marques, Garcia, and
  Sanchez}]{paper143}
Marques, A., Garcia, V., Sanchez, J., 2012{\natexlab{b}}. Two-level classifier
  ensembles for credit risk assessment. Expert Systems with Applications
  39~(12), 10916--10922.

\bibitem[{Marron(2007)}]{paper47}
Marron, D., 2007. 'lending by numbers': Credit scoring and the constitution of
  risk within american consumer credit. Economy and Society 36~(1), 103--133.

\bibitem[{Martens et~al.(2007)Martens, Baesens, Van, and Vanthienen}]{paper54}
Martens, D., Baesens, B., Van, Gestel, T., Vanthienen, J., 2007. Comprehensible
  credit scoring models using rule extraction from support vector machines.
  European Journal of Operational Research 183~(3), 1466--1476.

\bibitem[{Martens et~al.(2010)Martens, Van~Gestel, De~Backer, Haesen,
  Vanthienen, and Baesens}]{paper95}
Martens, D., Van~Gestel, T., De~Backer, M., Haesen, R., Vanthienen, J.,
  Baesens, B., 2010. Credit rating prediction using ant colony optimization.
  Journal of the Operational Research Society 61~(4), 561--573.

\bibitem[{Maznevski et~al.(2001)Maznevski, Kemp, Overstreet, and
  Crook}]{paper9}
Maznevski, M., Kemp, R., Overstreet, G., Crook, J., 2001. The power to borrow
  and lend: investigating the cultural context as part of the lending decision.
  Journal of the Operational Research Society 52~(9), 1045--1056.

\bibitem[{McDonald et~al.(2012)McDonald, Sturgess, Smith, Hawkins, and
  Huang}]{paper125}
McDonald, R., Sturgess, M., Smith, K., Hawkins, M., Huang, E., 2012.
  Non-linearity of scorecard log-odds. International Journal of Forecasting
  28~(1), 239--247.

\bibitem[{Mitchell(1997)}]{mitchell1997machine}
Mitchell, T.~M., 1997. Machine learning. 1997. Burr Ridge, IL: McGraw Hill 45.

\bibitem[{Mues et~al.(2004)Mues, Baesens, Files, and Vanthienen}]{paper23}
Mues, C., Baesens, B., Files, C., Vanthienen, J., 2004. Decision diagrams in
  machine learning: An empirical study on real-life credit-risk data. Expert
  Systems with Applications 27~(2), 257--264.

\bibitem[{Nieddu et~al.(2011)Nieddu, Manfredi, D'Acunto, and la}]{paper109}
Nieddu, L., Manfredi, G., D'Acunto, S., la, Regina, K., 2011. An optimal
  subclass detection method for credit scoring. World Academy of Science,
  Engineering and Technology 75, 349--354.

\bibitem[{Niklis et~al.(2014)Niklis, Doumpos, and Zopounidis}]{papern27}
Niklis, D., Doumpos, M., Zopounidis, C., 2014. Combining market and
  accounting-based models for credit scoring using a classification scheme
  based on support vector machines. Applied Mathematics and Computation 234,
  69--81.

\bibitem[{Nikolic et~al.(2013)Nikolic, Zarkic-Joksimovic, Stojanovski, and
  Joksimovic}]{papern51}
Nikolic, N., Zarkic-Joksimovic, N., Stojanovski, D., Joksimovic, I., 2013. The
  application of brute force logistic regression to corporate credit scoring
  models: Evidence from serbian financial statements. Expert Systems with
  Applications 40~(15), 5932--5944.

\bibitem[{Nurlybayeva and Balakayeva(2013)}]{papern62}
Nurlybayeva, K., Balakayeva, G., 2013. Algorithmic scoring models. Applied
  Mathematical Sciences 7~(9-12), 571--586.

\bibitem[{Nwulu and Oroja(2011)}]{paper116}
Nwulu, N., Oroja, S., 2011. A comparison of different soft computing models for
  credit scoring. World Academy of Science, Engineering and Technology 78,
  898--903.

\bibitem[{Nwulu et~al.(2012)Nwulu, Oroja, and Ilkan}]{paper147}
Nwulu, N., Oroja, S., Ilkan, M., 2012. A comparative analysis of machine
  learning techniques for credit scoring. Information 15~(10), 4129--4145.

\bibitem[{Ong et~al.(2005)Ong, Huang, and Tzeng}]{paper31}
Ong, C.-S., Huang, J.-J., Tzeng, G.-H., 2005. Building credit scoring models
  using genetic programming. Expert Systems with Applications 29~(1), 41--47.

\bibitem[{Paleologo et~al.(2010)Paleologo, Elisseeff, and Antonini}]{paper92}
Paleologo, G., Elisseeff, A., Antonini, G., 2010. Subagging for credit scoring
  models. European Journal of Operational Research 201~(2), 490--499.

\bibitem[{Pang(2005)}]{paper28}
Pang, S.-L., 2005. Study on credit scoring model and forecasting based on
  probabilistic neural network. Xitong Gongcheng Lilun yu Shijian/System
  Engineering Theory and Practice 25~(5), 43--48.

\bibitem[{Pavlidis et~al.(2012)Pavlidis, Tasoulis, Adams, and Hand}]{paper148}
Pavlidis, N., Tasoulis, D., Adams, N., Hand, D., 2012. Adaptive consumer credit
  classification. Journal of the Operational Research Society 63~(12),
  1645--1654.

\bibitem[{Ping and Yongheng(2011)}]{paper120}
Ping, Y., Yongheng, L., 2011. Neighborhood rough set and svm based hybrid
  credit scoring classifier. Expert Systems with Applications 38~(9),
  11300--11304.

\bibitem[{Ravi and Krishna(2014)}]{papern24}
Ravi, V., Krishna, M., 2014. A new online data imputation method based on
  general regression auto associative neural network. Neurocomputing 138,
  106--113.

\bibitem[{RBNZ(2013)}]{staff2013}
RBNZ, S., March 2013. Statement of principles: Bank registration and
  supervision financial stability. banking system handbook.

\bibitem[{Reddy and Ravi(2013)}]{papern61}
Reddy, K., Ravi, V., 2013. Differential evolution trained kernel principal
  component wnn and kernel binary quantile regression: Application to banking.
  Knowledge-Based Systems 39, 45--56.

\bibitem[{Rezac(2011)}]{paper111}
Rezac, M., 2011. Advanced empirical estimate of information value for credit
  scoring models. Acta Universitatis Agriculturae et Silviculturae Mendelianae
  Brunensis 59~(2), 267--274.

\bibitem[{{\v{R}}ez{\'a}{\v{c}} et~al.(2013){\v{R}}ez{\'a}{\v{c}}, Toma,
  et~al.}]{papern45}
{\v{R}}ez{\'a}{\v{c}}, M., Toma, L., et~al., 2013. Indeterminate values of
  target variable in development of credit scoring models. Acta Universitatis
  Agriculturae et Silviculturae Mendelianae Brunensis 61~(7), 2709--2716.

\bibitem[{Ripley(1996)}]{RIPLEY}
Ripley, B.~D., 1996. Pattern Recognition and Neural Networks. Cambridge
  University Press.

\bibitem[{Rohit et~al.(2013)Rohit, Kumar, and Kumar}]{Rohit13}
Rohit, V.~M., Kumar, S., Kumar, J., 2013. Basel {II} to basel {III} the way
  forward. Tech. rep., Infosys.

\bibitem[{Ruggieri et~al.(2010)Ruggieri, Pedreschi, and Turini}]{paper99}
Ruggieri, S., Pedreschi, D., Turini, F., 2010. Data mining for discrimination
  discovery. ACM Transactions on Knowledge Discovery from Data 4~(2).

\bibitem[{Saberi et~al.(2013)Saberi, Mirtalaie, Hussain, Azadeh, Hussain, and
  Ashjari}]{papern43}
Saberi, M., Mirtalaie, M., Hussain, F., Azadeh, A., Hussain, O., Ashjari, B.,
  2013. A granular computing-based approach to credit scoring modeling.
  Neurocomputing 122, 100--115.

\bibitem[{Sadatrasou et~al.(2015)Sadatrasou, Gholamian, and
  Shahanaghi}]{papern13}
Sadatrasou, S., Gholamian, M., Shahanaghi, K., 2015. An application of data
  mining classification and bi-level programming for optimal credit allocation.
  Decision Science Letters 4~(1), 35--50.

\bibitem[{Sadatrasoul et~al.(2015)Sadatrasoul, Gholamian, and
  Shahanaghi}]{papern17}
Sadatrasoul, S., Gholamian, M., Shahanaghi, K., 2015. Combination of feature
  selection and optimized fuzzy apriori rules: The case of credit scoring.
  International Arab Journal of Information Technology 12~(2), 138--145.

\bibitem[{Sarlin(2014)}]{papern38}
Sarlin, P., 2014. A weighted som for classifying data with instance-varying
  importance. International Journal of Machine Learning and Cybernetics 5~(1),
  101--110.

\bibitem[{Schapire(1990)}]{schapire1990}
Schapire, R.~E., 1990. The strength of weak learnability. Machine learning
  5~(2), 197--227.

\bibitem[{Setiono et~al.(2015)Setiono, Azcarraga, and Hayashi}]{papern1}
Setiono, R., Azcarraga, A., Hayashi, Y., 2015. Using sample selection to
  improve accuracy and simplicity of rules extracted from neural networks for
  credit scoring applications. International Journal of Computational
  Intelligence and Applications 14~(4).

\bibitem[{Setiono et~al.(2008)Setiono, Baesens, and Mues}]{paper57}
Setiono, R., Baesens, B., Mues, C., 2008. Recursive neural network rule
  extraction for data with mixed attributes. IEEE Transactions on Neural
  Networks 19~(2), 299--307.

\bibitem[{Setiono et~al.(2011)Setiono, Baesens, and Mues}]{paper118}
Setiono, R., Baesens, B., Mues, C., 2011. Rule extraction from minimal neural
  networks for credit card screening. International Journal of Neural Systems
  21~(4), 265--276.

\bibitem[{Sharma and Osei-Bryson(2009)}]{paper70}
Sharma, S., Osei-Bryson, K.-M., 2009. Framework for formal implementation of
  the business understanding phase of data mining projects. Expert Systems with
  Applications 36~(2 PART 2), 4114--4124.

\bibitem[{Shi et~al.(2013)Shi, Zhang, and Qiu}]{papern58}
Shi, J., Zhang, S.-Y., Qiu, L.-M., 2013. Credit scoring by feature-weighted
  support vector machines. Journal of Zhejiang University: Science C 14~(3),
  197--204.

\bibitem[{Shi(2009)}]{paper68}
Shi, Y., 2009. Current research trend: Information technology and decision
  making in 2008. International Journal of Information Technology and Decision
  Making 8~(1), 1--5.

\bibitem[{Shi(2010)}]{paper90}
Shi, Y., 2010. Multiple criteria optimization-based data mining methods and
  applications: A systematic survey. Knowledge and Information Systems 24~(3),
  369--391.

\bibitem[{Sinha and Zhao(2008)}]{paper65}
Sinha, A., Zhao, H., 2008. Incorporating domain knowledge into data mining
  classifiers: An application in indirect lending. Decision Support Systems
  46~(1), 287--299.

\bibitem[{So et~al.(2014)So, Thomas, Seow, and Mues}]{papern29}
So, M., Thomas, L., Seow, H.-V., Mues, C., 2014. Using a transactor/revolver
  scorecard to make credit and pricing decisions. Decision Support Systems
  59~(1), 143--151.

\bibitem[{Somol et~al.(2005)Somol, Baesens, Pudil, and Vanthienen}]{paper36}
Somol, P., Baesens, B., Pudil, P., Vanthienen, J., 2005. Filter- versus
  wrapper-based feature selection for credit scoring. International Journal of
  Intelligent Systems 20~(10), 985--999.

\bibitem[{Thomas(2010)}]{paper86}
Thomas, L., 2010. Consumer finance: Challenges for operational research.
  Journal of the Operational Research Society 61~(1), 41--52.

\bibitem[{Thomas et~al.(2002)Thomas, Edelman, and Crook}]{Thomas02}
Thomas, L.~C., Edelman, D., Crook, J., 2002. Credit Scoring and its
  applications. Monographs on mathematical modeling and computation. SIAM.

\bibitem[{Tomczak and Zie¸ba(2015)}]{papern12}
Tomczak, J., Zie¸ba, M., 2015. Classification restricted boltzmann machine for
  comprehensible credit scoring model. Expert Systems with Applications 42~(4),
  1789--1796.

\bibitem[{Tong et~al.(2012)Tong, Mues, and Thomas}]{paper136}
Tong, E., Mues, C., Thomas, L., 2012. Mixture cure models in credit scoring: If
  and when borrowers default. European Journal of Operational Research 218~(1),
  132--139.

\bibitem[{Tsai(2008)}]{paper63}
Tsai, C.-F., 2008. Financial decision support using neural networks and support
  vector machines. Expert Systems 25~(4), 380--393.

\bibitem[{Tsai(2009)}]{paper71}
Tsai, C.-F., 2009. Feature selection in bankruptcy prediction. Knowledge-Based
  Systems 22~(2), 120--127.

\bibitem[{Tsai(2014)}]{papern35}
Tsai, C.-F., 2014. Combining cluster analysis with classifier ensembles to
  predict financial distress. Information Fusion 16~(1), 46--58.

\bibitem[{Tsai et~al.(2014)Tsai, Hsu, and Yen}]{papern32}
Tsai, C.-F., Hsu, Y.-F., Yen, D., 2014. A comparative study of classifier
  ensembles for bankruptcy prediction. Applied Soft Computing Journal 24,
  977--984.

\bibitem[{Tsai and Wu(2008)}]{paper59}
Tsai, C.-F., Wu, J.-W., 2008. Using neural network ensembles for bankruptcy
  prediction and credit scoring. Expert Systems with Applications 34~(4),
  2639--2649.

\bibitem[{Van~Eyden(1995)}]{paper0}
Van~Eyden, R., 1995. Statistical modelling versus neural networks in financial
  decision making. Neural Network World 5~(1), 99--108.

\bibitem[{Van~Gestel et~al.(2007)Van~Gestel, Martens, Baesens, Feremans,
  Huysmans, and Vanthienen}]{paper50}
Van~Gestel, T., Martens, D., Baesens, B., Feremans, D., Huysmans, J.,
  Vanthienen, J., 2007. Forecasting and analyzing insurance companies' ratings.
  International Journal of Forecasting 23~(3), 513--529.

\bibitem[{Van~Gool et~al.(2012)Van~Gool, Verbeke, Sercu, and
  Baesens}]{paper133}
Van~Gool, J., Verbeke, W., Sercu, P., Baesens, B., 2012. Credit scoring for
  microfinance: Is it worth it? International Journal of Finance and Economics
  17~(2), 103--123.

\bibitem[{Vapnik(1998)}]{vapnik1998statistical}
Vapnik, V., 1998. Statistical learning theory.

\bibitem[{Verbraken et~al.(2014)Verbraken, Bravo, Weber, and
  Baesens}]{papern22}
Verbraken, T., Bravo, C., Weber, R., Baesens, B., 2014. Development and
  application of consumer credit scoring models using profit-based
  classification measures. European Journal of Operational Research 238~(2),
  505--513.

\bibitem[{Verstraeten and Van Den~Poel(2005)}]{paper32}
Verstraeten, G., Van Den~Poel, D., 2005. The impact of sample bias on consumer
  credit scoring performance and profitability. Journal of the Operational
  Research Society 56~(8), 981--992.

\bibitem[{Vukovic et~al.(2012)Vukovic, Delibasic, Uzelac, and
  Suknovic}]{paper139}
Vukovic, S., Delibasic, B., Uzelac, A., Suknovic, M., 2012. A case-based
  reasoning model that uses preference theory functions for credit scoring.
  Expert Systems with Applications 39~(9), 8389--8395.

\bibitem[{Waad et~al.(2013)Waad, Ghazi, and Mohamed}]{papern47}
Waad, B., Ghazi, B., Mohamed, L., 2013. A three-stage feature selection using
  quadratic programming for credit scoring. Applied Artificial Intelligence
  27~(8), 721--742.

\bibitem[{Wang et~al.(2011)Wang, Hao, Ma, and Jiang}]{paper108}
Wang, G., Hao, J., Ma, J., Jiang, H., 2011. A comparative assessment of
  ensemble learning for credit scoring. Expert Systems with Applications
  38~(1), 223--230.

\bibitem[{Wang et~al.(2012{\natexlab{a}})Wang, Ma, Huang, and Xu}]{paper127}
Wang, G., Ma, J., Huang, L., Xu, K., 2012{\natexlab{a}}. Two credit scoring
  models based on dual strategy ensemble trees. Knowledge-Based Systems 26,
  61--68.

\bibitem[{Wang et~al.(2012{\natexlab{b}})Wang, Hedar, Wang, and Ma}]{paper137}
Wang, J., Hedar, A.-R., Wang, S., Ma, J., 2012{\natexlab{b}}. Rough set and
  scatter search metaheuristic based feature selection for credit scoring.
  Expert Systems with Applications 39~(6), 6123--6128.

\bibitem[{Wang et~al.(2005)Wang, Wang, and Lai}]{paper37}
Wang, Y., Wang, S., Lai, K., 2005. A new fuzzy support vector machine to
  evaluate credit risk. IEEE Transactions on Fuzzy Systems 13~(6), 820--831.

\bibitem[{West(2000)}]{paper7}
West, D., 2000. Neural network credit scoring models. Computers and Operations
  Research 27~(11-12), 1131--1152.

\bibitem[{Wolpert(1992)}]{wolpert1992}
Wolpert, D.~H., 1992. Stacked generalization. Neural networks 5~(2), 241--259.

\bibitem[{Won et~al.(2012)Won, Kim, and Bae}]{paper150}
Won, C., Kim, J., Bae, J., 2012. Using genetic algorithm based knowledge
  refinement model for dividend policy forecasting. Expert Systems with
  Applications 39~(18), 13472--13479.

\bibitem[{Wu et~al.(2014)Wu, Olson, and Luo}]{papern21}
Wu, D., Olson, D., Luo, C., 2014. A decision support approach for accounts
  receivable risk management. IEEE Transactions on Systems, Man, and
  Cybernetics: Systems 44~(12), 1624--1632.

\bibitem[{Wu(2011)}]{paper119}
Wu, W.-W., 2011. Improving classification accuracy and causal knowledge for
  better credit decisions. International Journal of Neural Systems 21~(4),
  297--309.

\bibitem[{Xiao et~al.(2012)Xiao, Xie, He, and Jiang}]{paper129}
Xiao, J., Xie, L., He, C., Jiang, X., 2012. Dynamic classifier ensemble model
  for customer classification with imbalanced class distribution. Expert
  Systems with Applications 39~(3), 3668--3675.

\bibitem[{Xiao et~al.(2014)Xiao, Zhu, Teng, He, and Liu}]{papern40}
Xiao, J., Zhu, B., Teng, G., He, C., Liu, D., 2014. One-step dynamic classifier
  ensemble model for customer value segmentation with missing values.
  Mathematical Problems in Engineering 2014.

\bibitem[{Xiao et~al.(2006)Xiao, Zhao, and Fei}]{paper46}
Xiao, W., Zhao, Q., Fei, Q., 2006. A comparative study of data mining methods
  in consumer loans credit scoring management. Journal of Systems Science and
  Systems Engineering 15~(4), 419--435.

\bibitem[{Xiao and Fei(2006)}]{paper45}
Xiao, W.-B., Fei, Q., 2006. A study of personal credit scoring models on
  support vector machine with optimal choice of kernel function parameters.
  Xitong Gongcheng Lilun yu Shijian/System Engineering Theory and Practice
  26~(10), 73--79.

\bibitem[{Xiong et~al.(2013)Xiong, Wang, Mayers, and Monga}]{papern60}
Xiong, T., Wang, S., Mayers, A., Monga, E., 2013. Personal bankruptcy
  prediction by mining credit card data. Expert Systems with Applications
  40~(2), 665--676.

\bibitem[{Xu et~al.(2009)Xu, Zhou, and Wang}]{paper69}
Xu, X., Zhou, C., Wang, Z., 2009. Credit scoring algorithm based on link
  analysis ranking with support vector machine. Expert Systems with
  Applications 36~(2 PART 2), 2625--2632.

\bibitem[{Yang(2007)}]{paper56}
Yang, Y., 2007. Adaptive credit scoring with kernel learning methods. European
  Journal of Operational Research 183~(3), 1521--1536.

\bibitem[{Yang et~al.(2004)Yang, Wang, Bai, and Zhang}]{paper24}
Yang, Z., Wang, Y., Bai, Y., Zhang, X., 2004. Measuring scorecard performance.
  Lecture Notes in Computer Science (including subseries Lecture Notes in
  Artificial Intelligence and Lecture Notes in Bioinformatics) 3039, 900--906.

\bibitem[{Yap et~al.(2011)Yap, Ong, and Husain}]{paper122}
Yap, B., Ong, S., Husain, N., 2011. Using data mining to improve assessment of
  credit worthiness via credit scoring models. Expert Systems with Applications
  38~(10), 13274--13283.

\bibitem[{Yu and Li(2011)}]{paper123}
Yu, J.-L., Li, H., 2011. On performance of feature normalization in
  classification with distance-based case-based reasoning. Recent Patents on
  Computer Science 4~(3), 203--210.

\bibitem[{Zadeh(1965)}]{zadeh1965fuzzy}
Zadeh, L.~A., 1965. Fuzzy sets. Information and control 8~(3), 338--353.

\bibitem[{Zhang et~al.(2010)Zhang, Zhou, Leung, and Zheng}]{paper106}
Zhang, D., Zhou, X., Leung, S., Zheng, J., 2010. Vertical bagging decision
  trees model for credit scoring. Expert Systems with Applications 37~(12),
  7838--7843.

\bibitem[{Zhang et~al.(2014)Zhang, Gao, and Shi}]{papern25}
Zhang, Z., Gao, G., Shi, Y., 2014. Credit risk evaluation using multi-criteria
  optimization classifier with kernel, fuzzification and penalty factors.
  European Journal of Operational Research 237~(1), 335--348.

\bibitem[{Zhao et~al.(2015)Zhao, Xu, Kang, Kabir, Liu, and Wasinger}]{papern5}
Zhao, Z., Xu, S., Kang, B., Kabir, M., Liu, Y., Wasinger, R., 2015.
  Investigation and improvement of multi-layer perception neural networks for
  credit scoring. Expert Systems with Applications 42~(7), 3508--3516.

\bibitem[{Zhou et~al.(2009)Zhou, Lai, and Yen}]{paper82}
Zhou, L., Lai, K., Yen, J., 2009. Credit scoring models with auc maximization
  based on weighted svm. International Journal of Information Technology and
  Decision Making 8~(4), 677--696.

\bibitem[{Zhou et~al.(2010)Zhou, Lai, and Yu}]{paper87}
Zhou, L., Lai, K., Yu, L., 2010. Least squares support vector machines ensemble
  models for credit scoring. Expert Systems with Applications 37~(1), 127--133.

\bibitem[{Zhu et~al.(2002)Zhu, Beling, and Overstreet}]{paper16}
Zhu, H., Beling, P., Overstreet, G., 2002. A bayesian framework for the
  combination of classifier outputs. Journal of the Operational Research
  Society 53~(7), 719--727.

\bibitem[{Zhu and Hu(2013)}]{papern57}
Zhu, P., Hu, Q., 2013. Rule extraction from support vector machines based on
  consistent region covering reduction. Knowledge-Based Systems 42, 1--8.

\bibitem[{Zhu et~al.(2013)Zhu, Li, Wu, Wang, and Liang}]{papern46}
Zhu, X., Li, J., Wu, D., Wang, H., Liang, C., 2013. Balancing accuracy,
  complexity and interpretability in consumer credit decision making: A
  c-topsis classification approach. Knowledge-Based Systems 52, 258--267.

\bibitem[{Ziari et~al.(1997)Ziari, Leatham, and Ellinger}]{paper2}
Ziari, H., Leatham, D., Ellinger, P., 1997. Development of statistical
  discriminant mathematical programming model via resampling estimation
  techniques. American Journal of Agricultural Economics 79~(4), 1352--1362.

\bibitem[{Zweig and Campbell(1993)}]{Zweig}
Zweig, M.~H., Campbell, G., 1993. Receiver-operating characteristic {(ROC)}
  plots: a fundamental evaluation tool in clinical medicine. Clinical Chemistry
  29, 561--577.

\end{thebibliography}
\end{document}